\documentclass[aps,prx,longbibliography,amsmath,amssymb,12pt]{revtex4-1}
\usepackage{graphicx}
\usepackage{amsmath}
\usepackage{amssymb}
\usepackage{amstext}
\usepackage{array}
\usepackage{amsfonts}
\usepackage{upgreek}
\usepackage{txfonts}
\usepackage{enumerate}
\usepackage{comment}

\usepackage{hyperref}
\usepackage{xcolor}
\definecolor{dark-red}{rgb}{0.4,0.15,0.15}
\definecolor{dark-blue}{rgb}{0.15,0.15,0.4}
\definecolor{medium-blue}{rgb}{0,0,0.5}
\hypersetup{
colorlinks, linkcolor={dark-blue},
citecolor={dark-blue}, urlcolor={medium-blue}
}

\begin{document}

\title{Universality in antiferromagnetic strange metals}

\author{Stefan A.\ Maier$^1$}
\email{Stefan.Maier@uni-koeln.de}

\author{Philipp Strack$^{1}$}
\email{strack@thp.uni-koeln.de}
\homepage{http://www.thp.uni-koeln.de/~strack}

\affiliation{$^{1}$Institut f\"ur Theoretische Physik, Universit\"at zu K\"oln, D-50937 Cologne, Germany}

\date{\today}

\begin{abstract}
We propose a theory of metals at the spin-density wave 
quantum critical point in spatial dimension $d=2$. We provide a first estimate of 
the full set of critical exponents
(dynamical exponent $z=2.13$, correlation length $\nu =1.02$, 
spin susceptibility $\gamma = 0.96$, 
electronic non-Fermi liquid $\eta^f_\tau = 0.53$, spin-wave Landau damping
$\eta^b_\tau = 1.06$), which determine the universal power-laws in thermodynamics 
and response functions in the quantum-critical regime relevant for experiments in 
heavy-fermion systems and iron pnictides. We present approximate numerical and 
analytical solutions of Polchinski-Wetterich type flow equations with 
soft frequency regulators for an effective action of electrons coupled to 
spin-wave bosons. Performing the renormalization group in frequency- 
instead of momentum-space allows to track changes of the Fermi surface shape 
and to capture Landau damping during the flow. The technique 
is easily generalizable from models retaining only patches of the 
Fermi surface to full, compact Fermi surfaces.
%
%
\end{abstract}

\maketitle
\tableofcontents

\section{Introduction}
The universality hypothesis in physics states that seemingly 
disparate materials, with different lattice structure, 
different chemical composition, and at different overall 
temperatures can obey, when undergoing a phase transition, the same universal laws
effective at long distances \cite{stanley99,goldenfeld_book}. 
Rather than the complex 
microscopic details, what matters for a material 
to belong to a universality class described by the same set of critical exponents 
are the small number of independent parameters associated with symmetries 
such as spin-rotation O(3) invariance (a three-parameter group) in magnets.
It is a major success of the renormalization group (RG) \cite{wilson75} 
to reduce the complexity of many microscopic details during the 
flow from high to low energies along a cutoff parameter $\Lambda$ 
and distill out the relevant features and universality class of a given model in the 
asymptotic limit $\Lambda\rightarrow 0$, when the flow is attracted by 
a scale-invariant fixed-point.

For metals, that is, solid state materials with freely moving electrons and 
a Fermi surface in momentum space, undergoing phase transitions at 
zero temperature, not much is known at present about the emergence
of universal quantum critical phenomena from different 
microscopic models. What makes the problem so difficult is the interplay of two types 
of low-energy singularities: the electronic Green's function becomes unbounded 
on the entire Fermi surface with co-dimension $d-1$ and fluctuations of the (bosonic) order parameter 
are infrared divergent at the origin $\omega=0$, $\mathbf{q}^2 = 0$ of frequency- and momentum space 
at the quantum critical point. Answers to this question will 
result in more precise predictions and long-sought explanations 
for findings in experiments in at least three different material classes:
the heavy-fermion compounds \cite{stewart-heavy-ferm,loehneysen-review}, 
iron pnictides \cite{matsuda14}, and the cuprates \cite{subir_cuprates}. 
The case of two space dimensions is of particular interest.

\begin{figure}
	\includegraphics[width=100mm]{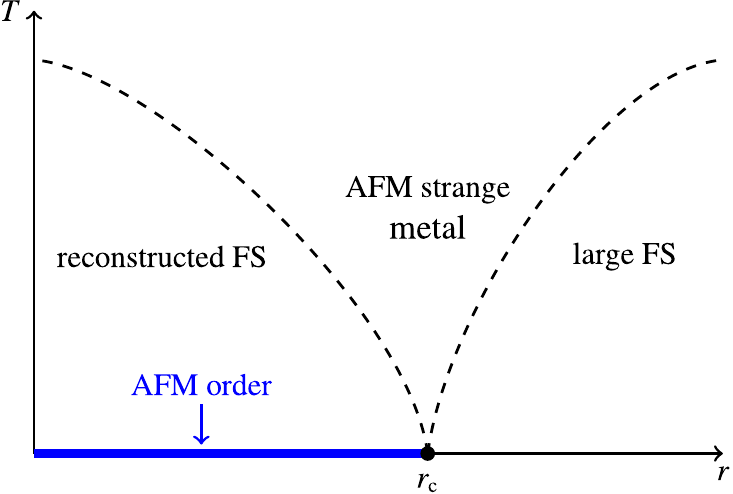}
	\caption{Generic phase diagram of the spin-density wave 
	quantum critical point in two-dimensional metals at $r_c$ in the 
	temperature ($T$) and control parameter ($r$) plane. With zero-temperature 
	antiferromagnetic order (AFM order), the large Fermi surface in Fig.~\ref{fig:fs-hs} reconstructs
	with gaps opening at the hot spots. This paper establishes the universal 
	critical physics in $d=2$ which will enable systematic renormalization of competing 
	orders such as superconductivity and charge density wave order.}
	\label{fig:phase-diag}
\end{figure}

\begin{figure}
	\includegraphics[width=\linewidth]{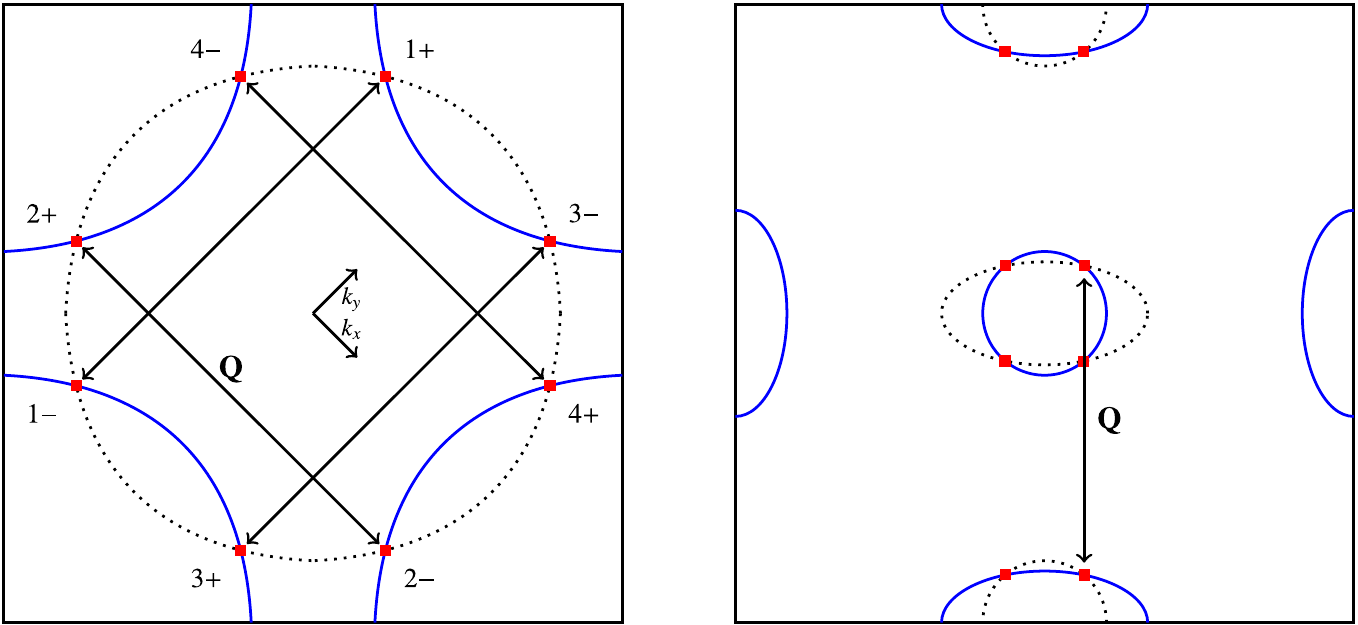}
	\caption{Prototypical Fermi surfaces for commensurate antiferromagnetism in metals. 
	Large Fermi-surface setup (as e.g. in the cuprates \cite{ACS03}, left) and with smaller pockets (as e.g. in 
	the iron pnictides, right).
		The Fermi surfaces are depicted as blue solid lines in the first BZ,
		and the \emph{hot spots} of the dispersion as red dots.
		In this paper, we will focus on the large Fermi surface setup (left), where there are
         four pairs of hot spots on the Fermi surface connected by the commensurate ordering vector $\mathbf{Q}= (\pi,\pi)$.
	 They are labeled by the pair index $n = 1,2,3,4 $ and by the Fermi-surface index $ m = \pm $
	 (Note that the momentum coordinates $k_x$ and $k_y$ used in this work correspond to the diagonal of the BZ.).
	 In the pnictides \cite{matsuda14}, two pairs of hot spots for the magnetic stripe ordering are connected by  		
	$ \mathbf{Q} = (0,\pi) $. The critical properties of the bi-directional antiferromagnet 
	in the small pockets case are expected to be similar to the large Fermi surface.}
	\label{fig:fs-hs}
\end{figure}

Serious evidence for the existence of 
fixed-point behavior in a two-dimensional metal 
has recently been presented by Sur and Lee \cite{sur14} 
for a large Fermi surface at the onset of 
antiferromagnetism with 
the general phase diagram shown in Fig.~\ref{fig:phase-diag}
\footnote{The problem of the nematic metal is also open again 
due to surprises appearing at higher loop order in diagrammatic studies 
(see Ref.~\onlinecite{holder15} and references therein) and in 
numerical studies \cite{lederer15}.}.
Sur and Lee devised an embedding of the Fermi surface 
(Fig.~\ref{fig:fs-hs}) into higher-dimensional (frequency) space 
that enabled an $\epsilon = 3 - d$ expansion by effectively turning the metal 
into a semimetal with vanishing density of states at the Fermi level. 
Extrapolating the results of a 
field-theoretical renormalization group to $\epsilon=1$, a strange metallic 
fixed-point theory with dynamical exponent $z>1$ was found and interpreted as an 
interactionless and dispersionless state of a local boson coupled to four sets of dimensionally 
reduced, counter-propagating one-dimensional electrons. 
We will below offer an alternative interpretation of the 
antiferromagnetic strange metal in exactly two dimensions.

\subsection{Key results}
\label{sec:key-results}

In this paper, we test, and for the first time, explicitly confirm the universality hypothesis for metals 
directly in $d=2$ focusing on the above-mentioned quantum critical antiferromagnet.
To achieve this, we solve a simple set of renormalization group flow equations 
derived from the scale-dependent effective action due to Polchinski \cite{polchinski84}
and Wetterich \cite{wetterich93,metzner_review}.

\begin{figure}
	\includegraphics[width=.5\linewidth]{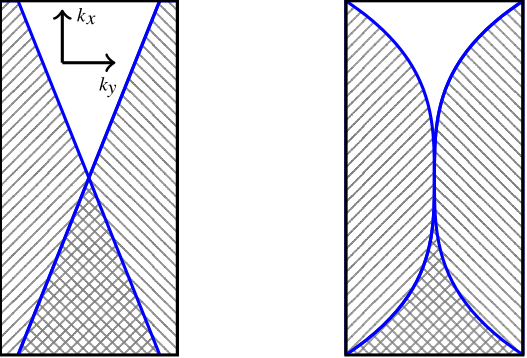}
	\caption{Intersection of the two Fermi surfaces of one hot-spot pair, with the two hot spots on top of one another.
	 The shading indicates the sign of the fermionic dispersion, i.e.\ the dispersion has a negative sign in the shaded areas.
	 In the left part, the situation at high energy scales is depicted, where the Fermi surfaces intersect with a finite angle.
	 The right part illustrates the situation in the deep IR, i.e.\ at $\Lambda=0$, when the angle $\varv$ 
	 between the Fermi surfaces logarithmically flows to zero as shown in Fig.~\ref{fig:flow-v}.
	  The Fermi surfaces are then locally nested at a single point independent of 
	  the bare angle, provided it is not nested initially.}
	\label{fig:fs-angle}
\end{figure}

It is known for some time \cite{abanov00}
that spin-wave fluctuations locally deform the Fermi surface (Fig.~\ref{fig:fs-angle})
but it is precisely this feature that destroyed the scaling within the 
field-theoretical RG analysis of Metlitski and Sachdev \cite{MMSS10b} 
leading to anomalous logarithmic corrections and divergences 
in various critical exponents.

Here, we overcome this problem by devising soft frequency regulators generalizing the $\Omega$-flow scheme
\cite{giering-nfl}, that allow the shape of the 
Fermi surface to change during the flow, to dynamically adaptive 
form \cite{dynRG};
this performs well with correlations that build up during the flow 
and allows penetration of the quantum-critical regime. In particular, 
as we argue in Sec.~\ref{subsec:weak}, the strange metal 
fixed-point we find is accessible from an initially weakly coupled 
model and therefore amendable to the combined vertex and phase-space 
expansion inherent in the functional RG hierarchy
with $\Lambda$-dependent adaptation \cite{metzner_review,dynRG}.

\begin{figure}
 \includegraphics[width=50mm]
 {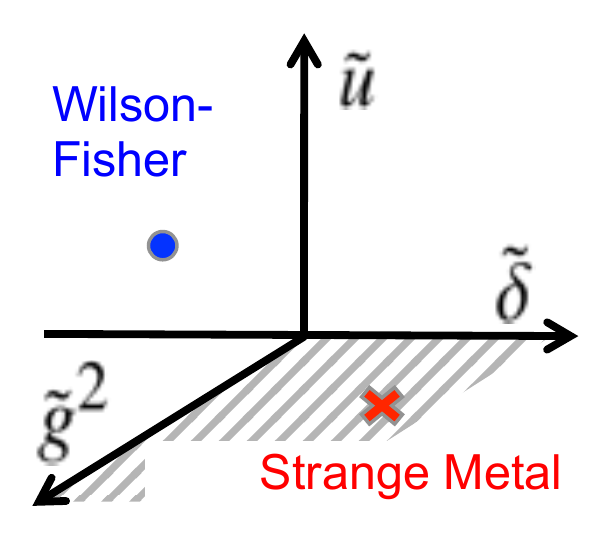}
 \caption{Coupling the Wilson-Fisher fixed-point of  the $O(3)$ model to eight patches of Fermi surface 
 results in a $d=2$ strange metal fixed point (with $\tilde{u}=0$ and $\tilde{g}^2\neq 0$) with novel values of 
 critical exponents.}
 \label{fig:sm-fixed}
\end{figure}

Using this technique, we are able to present a rather complete, analytical and 
numerical ``RG tomography'' of the problem across the entire range of scales, 
from high energies to the low-energy asymptotics $\Lambda\rightarrow 0$. 
We begin by presenting the universal set of $\beta$-functions
in $d=2$, 
which (i) are completely $\Lambda$/scale-independent, (ii) contain only 
``dimensionless'' variables free of logarithmic corrections, and (iii) 
enable access to the $\varv \rightarrow 0$ limit of asymptotically 
nested hot spots (cf.~ Fig.~\ref{fig:fs-angle}):
  \begin{align}
  	  \label{eq:beta-deltat}
  	  \beta_{\tilde{\delta}} &= 
	  - \left( 2 - \eta^b_\tau \right) \tilde{\delta}
- \frac{5}{\pi}  \tilde{u} \left( 1 - \eta^b_\tau / 2 \right) \left(1 + \tilde{\delta} \right)^{-1/2}
+3 \tilde{g}^2 \\
	  \label{eq:beta-ut}
	  \beta_{\tilde{u}}  &= 
	  \tilde{u} \left(\eta^b_\tau -1 \right) + \frac{11 \tilde{u}^2}{ 2 \pi}
             \left( 1 - \eta^b_\tau / 2 \right) \left(1 + \tilde{\delta} \right)^{-3/2} \\
               \label{eq:beta-gt}
	  \beta_{\tilde{g}}  &= 
	  \left[-1 + \eta^b_\tau-2 \tilde{g}^2 T_g (\varv\to 0,\varw,\eta^b_\tau,\tilde{\delta})   \right] \frac{\tilde{g}}{2} \\
	   \label{eq:beta-w}
	  \beta_\varw 
	  &= \left[ 6 \tilde{g}^2 T^f_\tau (\varv\to 0,\varw,\eta^b_\tau,\tilde{\delta}) 
	  - \eta^b_\tau  \right] \frac{\varw}{2} 
	  \\
	  \label{eq:beta-eta}
	   \eta^b_\tau &= \tilde{g}^2 T^b_\tau = \frac{\tilde{g}^2}{4}
  \end{align}
where the dimensionless threshold functions $T_g$, $T^f_\tau$, and $T^b_\tau$ 
are defined in Sec.~\ref{sec:orig-screl}. The first two equations bear resemblance to the $O(3)$ Wilson-Fisher fixed-point 
of the bosonic three-component order parameter $\vec{\phi}$ with $\tilde{\delta}$ the mass and 
$\tilde{u}$ the quartic coupling. Crucially here, these equations are coupled to 
eight patches of the Fermi surface via the coupling $\tilde{g}$ and the main 
contribution of an anomalous dimension of $\vec{\phi}$ comes from electrons 
close to the Fermi surface via Eq.~(\ref{eq:beta-eta}). Additional factors 
of the anomalous dimension $\eta^b_\tau$ appear due to dynamical adaption \cite{dynRG}.
Moreover there is a ``velocity ratio'' $\varw$ that flows to a universal fixed-point value.
We illustrate the novel $d=2$ fixed-point of Eqs.~(\ref{eq:beta-deltat}-\ref{eq:beta-eta}) in 
Fig.~\ref{fig:sm-fixed} in comparison to the Wilson-Fisher fixed point of the order-parameter 
theory only. 
The full set of critical exponents can be extracted from 
Eqs.~(\ref{eq:beta-deltat}-\ref{eq:beta-eta}) (for details see Subsec.~\ref{sec:flow-mass}).
The Landau damping exponent $\eta_\tau^{b}$ of the spin waves, 
electronic non-Fermi liquid exponent $\eta_\tau^{f}$, and the power-law divergence 
of the electron-boson vertex $ \eta_g$ attain the fixed-point values:
  \begin{equation}
	  \eta^{b\ast}_\tau = 1.062 \, , \quad \eta^{f\ast}_\tau = 0.530 \, , \quad \eta_g^\ast = 0.03\;.
	  \label{eq:values-etas}
  \end{equation}
The $\eta_\tau^{f}$ appears implicitly in Eqs.~(\ref{eq:beta-deltat}-\ref{eq:beta-eta})
via the scaling relations: 
\begin{align} 
\label{eq:screl-g-abt}
 \eta^b_\tau &= 1 + 2 \eta_g\\
 \label{eq:fermbos-screl}
 \eta^b_\tau /2 &=  \eta^f_\tau + \eta^f_y\;,
\end{align}
with $\eta^{f\ast}_y = 0$ but the electronic dispersions in $y$- (and $x$)-direction 
are renormalized logarithmically.

Further we get for the dynamical exponent
\begin{align}
 z_f = \frac{1- \eta^{f\ast}_y}{1-\eta^{f\ast}_\tau} = 2.13 
= z_b = \frac{1-\eta^{b\ast}_{xy}/2}{1-\eta^{b\ast}_\tau/2}\;.
\label{eq:dynamical}
\end{align}
with $\eta^{b\ast}_{xy}=0$ and $\eta^{f\ast}_y = - \eta^{f\ast}_x = 0$.
Note that an additive definition of the dynamical exponent 
(also used in Ref.~\onlinecite{sur14} and Fig.~\ref{fig:zs}) is
$z^\ast_f = 1 + \eta^{f\ast}_\tau - \eta^{f\ast}_y
= 1.53=z^\ast_b = 1 + \frac{\eta^{b\ast}_\tau - \eta^{b\ast}_{xy}}{2}$.

\begin{figure}[t]
 \includegraphics[width=\linewidth]{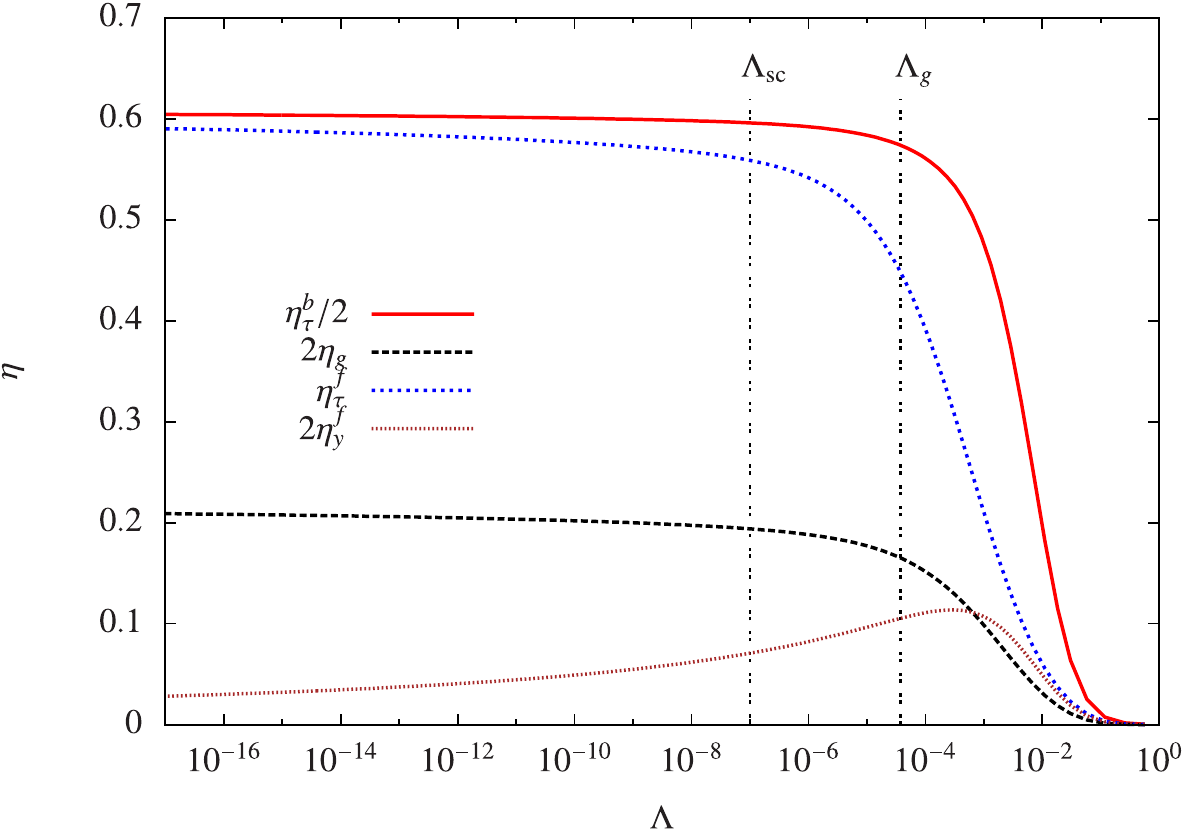}
 \caption{Flow (with pinned mass $\delta_b=0\;\forall \Lambda$) of the anomalous exponents from high energies (large $\Lambda$, right) 
 to low energies (small $\Lambda$, left) indicative of power-law scaling in various 
 physical observables at the strange metal fixed-point, as summarized in Sec.~\ref{sec:key-results}.
 The RG flow depicted is pinned to criticality (see Sec.~\ref{sec:numerics}) for 
 purely technical reasons of numerical accuracy. This shifts the IR values
 of the exponents compared to those of the massive flow quoted in 
 Sec.~\ref{sec:key-results} and derived in Sec.~\ref{sec:flow-mass}.
 Initial values for the flow are $\mathcal{A}^b_{\tau,\mathrm{bare}} = 1.0$, 
 $\mathcal{A}^b_{xy,\mathrm{bare}} = 0.5$, $ g_\mathrm{bare}^2 = 0.007$, $ \varv_\mathrm{bare} =0.3$, 
 $A^f_{\tau,\mathrm{bare}} = A^f_{y,\mathrm{bare}} = 1.0 $, $ \Lambda_\mathrm{UV} = 0.1$. 
 The double-dotted vertical lines mark the crossover scales between the different regimes of the 
 RG flow explained in Sec.~\ref{sec:regimes}.
 }
 \label{fig:etas}
\end{figure}
Based on computations in Sec.~\ref{sec:flow-mass}, we can report
values for the correlation length ($\xi$) and spin susceptibility 
($\chi$) exponents when approaching the critical point as a function 
of control parameter
\begin{align}
\xi &  \sim  \frac{1}{(r - r_c)^\nu}\;,\;\;\; \nu = 1.019\nonumber\\
\chi & \sim \frac{1}{(r- r_c)^\gamma }\;,\;\;\; \gamma = 0.956\;.
\label{eq:crit-exp}
\end{align}

\begin{figure}
 \includegraphics[width=\linewidth]{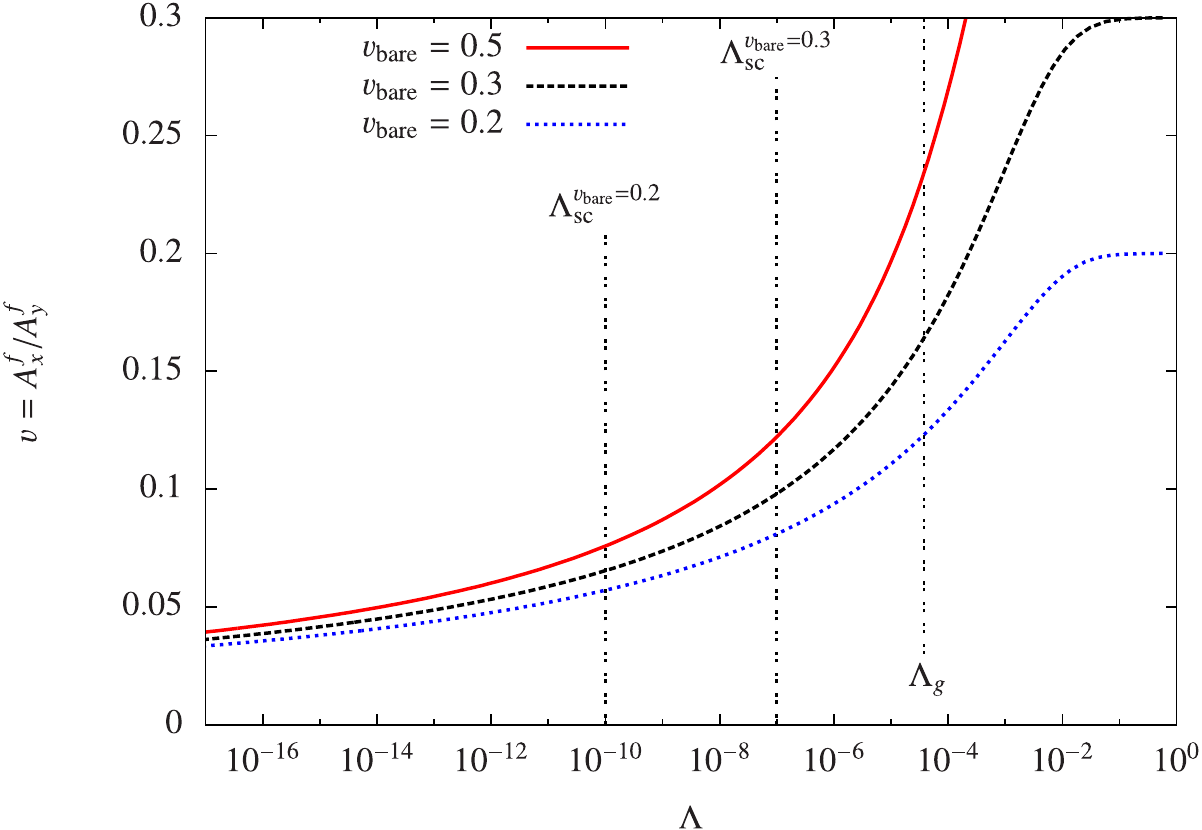}
 \caption{Explicit evidence of universality in the antiferromagnetic strange metal at low energies $\Lambda\rightarrow 0$: Flow (for pinned mass $\delta_b = 0\; \forall \Lambda$) of the Fermi surface angle $ \varv \equiv A^f_x / A^f_y $ for $\mathcal{A}^b_{\tau,\mathrm{bare}} = 1.0$, $\mathcal{A}^b_{xy,\mathrm{bare}} = 0.5$, $ g_\mathrm{bare}^2 = 0.007$, $ A^f_{\tau,\mathrm{bare}} = A^f_{y,\mathrm{bare}} = 1.0 $, $ \Lambda_\mathrm{UV} = 0.1$. The three curves correspond to different values of initial angles between the Fermi surfaces $ \varv_\mathrm{bare}$ and exhibit universal logarithmic flow towards zero $\varv \rightarrow 0$. For other universal features, see Figs.~\ref{fig:g-rat},\ref{fig:w-rat}.}
 \label{fig:flow-v}
\end{figure}

The analytic findings of the fixed-point solution are confirmed 
in an explicit numerical solution across all $\Lambda$ and 
lead to the following physical predictions. 
The anomalous power-law scaling of the effective action 
(see Fig.~\ref{fig:etas} and the RG flows in Sec.~\ref{sec:numerics})
over many orders of magnitude leads to pronounced non-Fermi liquid 
behavior for the electrons at the hot spots. The physical electron Green's function for momenta 
in the vicinity of the hot spots ($n$=1 without loss of generality) attains the form
\begin{equation}
\langle \bar{\psi}(\omega,\mathbf{k}) \, \psi(\omega,\mathbf{k}) \rangle
\sim \frac{1}{\omega^{1-\eta^f_\tau} + A^f_x k_x \pm A^f_y k_y} \, , \;\;\;\; 
\eta^{f\ast}_\tau = 0.53
\label{eq:elec}
\end{equation}
with only logarithmic corrections to the momentum dispersion via $A^f_x$, $A^f_y$. 
The Fermi surface angle $\varv = \frac{A^f_x}{A^f_y}$ plotted in 
Figs.~\ref{fig:fs-angle},\ref{fig:flow-v} flows to zero logarithmically 
as analyzed in detail below in Sec.~\ref{sec:numerics}, Fig.~\ref{fig:eta-v}. As a 
consequence of Eq.~(\ref{eq:elec}), one may define a critical contribution to the 
tunneling density of states 
%
$
	N_\mathrm{hs} (\omega) \equiv \frac{1}{4 \pi^3} \int_{-\Lambda_\mathrm{UV}}^{+\Lambda_\mathrm{UV}} \! d k_x \,  \int_{-\Lambda_\mathrm{UV}}^{+\Lambda_\mathrm{UV}} \! d k_y \, \left| \operatorname{Im} G^\mathrm{ret} (\omega,\mathbf{k}) \right|
	\sim \omega^{0.47} $.
%
	Note that the cold regions along the Fermi surface outside the UV cutoff $\Lambda_\mathrm{UV}$ are however 
expected to provide the dominant, constant Fermi-liquid contributions to the local density of states.

For the frequency- and momentum resolved spin susceptibility around $\mathbf{Q}=(\pi,\pi)$ 
as measured in neutron scattering, we predict
\begin{align}
\langle \vec{\phi}(-\omega, -\mathbf{q}) \, \vec{\phi}(\omega, \mathbf{q}) \rangle 
\sim 
\frac{1}{\omega^{2-\eta^b_\tau} + \mathbf{q}^{2-\eta^b_{xy}}} = \frac{1}{\omega^{0.94} + \mathbf{q}^2} 
\label{eq:spin_suscept}
\end{align}
quoting the values $\eta^{b\ast}_\tau = 1.06$ and $\eta^{b\ast}_{xy} =0$ from Eq.~(\ref{eq:values-etas})
and Sec.~\ref{sec:flow-mass}.
At the critical wavevector $ \mathbf{q} =0 $, one obtains for the spin susceptibility 
$ \chi_\mathrm{s} \sim \omega^{- \alpha} $ with $ \alpha = 0.94 $.
Measurements of this susceptibility for CeCu$_{5.9}$Au$_{0.1}$ \cite{schroeder-nature-2000} give 
$\alpha = 0.75$ with however a weak momentum dependence.

We note here that both, the physical electron propagator in Eq.~(\ref{eq:elec}) and the 
physical spin fluctuation propagator in Eq.~(\ref{eq:spin_suscept}) remain fully two dimensional, 
which leads to the hyperscaling property found in Ref.~\onlinecite{patel15}. 
The fixed-point we find here remains also interacting (as seen from a small, finite anomalous exponent $\eta_g$ 
of the electron-spin wave coupling $g$ in Fig.~\ref{fig:etas}). 
Both of the above properties are in contrast to the interpretation offered by Sur and Lee \cite{sur14}
although some of their critical exponents are qualitatively similar to ours, as we outline 
in Sec.~\ref{subsec:surlee}.

For thermodynamic measurements \cite{stewart-heavy-ferm,loehneysen-review}, 
we predict the electrons from the hot spots to obey
%
\begin{equation} \label{eq:c-el}
C_{e} /T \sim  T^{\frac{d-\theta}{z^\ast_f}-1} = T^{-0.061}\;,
\end{equation}
which depends on spatial dimensionality ($d=2$), the hyperscaling-violation exponent 
($\theta =0$ as hyperscaling is preserved), and the electronic dynamical exponent
Eq.~(\ref{eq:dynamical}). As before, 
the hot spot contribution Eq.~(\ref{eq:c-el}) is expected to be supplemented by 
contributions from cold, or lukewarm contributions from the rest of the Fermi surface.
From the dynamical exponent and the correlation length exponent
reported above
follows 
the scaling of the Gr\"{u}neisen ratio as a function of temperature \cite{garst-rosch-grueneisen}
\begin{equation} \label{eq:grueneisen}
\Gamma_\mathrm{GE}\sim T^{-1/(\nu\, z^\ast_f)} = T^{-0.46}\;,
\end{equation}
broadly in the same range as observed experimentally 
with exponents between $ - 0.7 $ and $ -1.0 $ for different heavy-fermion compounds 
\cite{kuechler-grueneisen}.

Finally, we confirm the universality hypothesis of the fixed-point as 
the energies are lowered toward $\Lambda \rightarrow 0$. 
Together with Figs.~\ref{fig:g-rat} and \ref{fig:w-rat}, Fig.~\ref{fig:flow-v} displays explicit evidence that the same 
scaling with logarithmically vanishing angle $\varv$ 
between the Fermi surfaces is attained independent of the initial 
value of the angle between the Fermi surfaces/details of the band structure. 
Taken at face value, this implies that the infrared low-energy exponents of the
quantum critical antiferromagnetic metal should be same across the various material categories, 
pnictides, cuprates, or heavy fermions. The fact that they are {\it not}, according to most 
measurements, implies that other and/or additional forces on top of 
(clean) antiferromagnetic quantum criticality are at the root of strange metallic behavior.
Also, at higher scales, above the 
crossover scale $\Lambda_\mathrm{sc}$ non-universal material specific features 
can lead to different exponents between the materials.

The main body of the paper is organized as follows: in Sec.~\ref{sec:model}, 
we introduce the spin-fermion model and describe the eight parameters for which we compute 
the RG flows. In Sec.~\ref{sec:RG}, we introduce the soft frequency regulators 
and dynamical adaptation technique.
In Sec.~\ref{sec:numerics}, we present a numerical solution 
of the coupled flow equations pinned to zero bosonic mass and and discuss 
various crossover scales. In Sec.~\ref{sec:orig-screl}, 
we give analytical expressions for the universal $\beta$-functions at the 
strange metal fixed-point and derive the scaling relations found in the numerics.
In that section, we explain how Landau damping appears naturally under the RG with soft frequency cutoffs.
In Sec.~\ref{sec:comparison}, we compare and discuss differences to important previous work on the 
spin-density wave quantum-critical point in two-dimensional metals.  

\section{Spin-fermion model for the antiferromagnetic strange metal in $d=2$}
\label{sec:model}

A minimal model to capture the critical behavior of itinerant electrons in a two-dimensional 
metal interacting with self-generated, commensurate $\mathbf{Q}=(\pi,\pi)$ spin waves 
should contain four terms, 
$\Gamma^\Lambda = \Gamma^\Lambda_f + \Gamma^\Lambda_b + \Gamma^\Lambda_\mathrm{int} + \Gamma^\Lambda_{\phi^4}$,
and a total of eight parameters, which we now describe. 
The Fermi surface set-up is depicted in Fig.~\ref{fig:fs-hs},
labeling and choice of coordinate axis follow Sur and Lee \cite{sur14}.
These parameters depend on 
the running energy scale $\Lambda$ and we will compute their flow 
from their bare values at high energies $\Lambda_\mathrm{UV}$ 
to their renormalized values at low energies $\Lambda \to 0$.

The propagator for electrons close to the hot spots is given by
\begin{align}
\Gamma_f^\Lambda[\bar{\psi},\psi]
=
\sum_{n=1}^4 \sum_{m=\pm} \sum_{\sigma = \uparrow,\downarrow}
\int \frac{ d^3 k}{(2\pi)^3} \,
\bar{\psi}^{(m)}_{n,\sigma}(k) 
\left[
A^f_{\tau} i k_\tau + e^{(m)}_n(\mathbf{k},\Lambda)
\right]
\psi^{(m)}_{n,\sigma}(k)\;,
\label{eq:ferm_prop}
\end{align}
where the three momentum is defined as $k = (k_\tau,\mathbf{k}) = (k_\tau, k_x, k_y)$, 
contains three parameters: $A^f_\tau$ for the frequency dependence 
and $A^f_x$, $A^f_y$ for the dispersions
\begin{align}
e^{(\pm)}_1(\mathbf{k},\Lambda)& = -e^{(\pm)}_3(\mathbf{k},\Lambda)
= A^f_{x} k_x \pm A^f_{y} k_y
\nonumber\\
e^{(\pm)}_2(\mathbf{k},\Lambda) &= -e^{(\pm)}_4(\mathbf{k},\Lambda)
=\mp  A^f_{24,x} k_x + A^f_{24,y} k_y\;,
\label{eq:dispersions}
\end{align}
where furthermore $A^f_{x} =A^f_{24,y}$ and $A^f_{y} = A^f_{24,x}$ due to the point group symmetries.
The numerical values of $A^f_x$ and $A^f_y$-factors determine the position of the 
Fermi surface $e^{(m)}_n(\mathbf{k}_\mathrm{FS},\Lambda)=0$. 
We have suppressed the $\Lambda$-dependence in the $A$-factors for 
notational clarity. 
Eq.~(\ref{eq:dispersions}) restricts to terms linear in momentum. 
However, Fermi surface curvature terms $\propto k_x^3, k_y^3$ play a role 
for electrons away from the hot spots and we will capture some of these 
effects at initial stages of the flow $\Lambda \lesssim \Lambda_\mathrm{UV}$ through 
explicit use of a UV-cutoff $\Lambda_\mathrm{UV}$.

The bosonic propagator will be parametrized by three parameters 
$\mathcal{A}^b_\tau$, $\mathcal{A}^b_{xy}$ and $\delta_b$:
\begin{align}
\Gamma_b^\Lambda[\vec{\phi}]
=\frac{1}{2} 
\int \! \frac{ d^3 q}{(2\pi)^3}  \,
\left[ \mathcal{A}^b_{\tau} q_\tau^2 
+\mathcal{A}^b_{xy}\left( q_x^2 +  q_y^2\right)
+ \delta_b
\right]
\vec{\phi}(-q) \cdot \vec{\phi}(q)\;.
\label{eq:bose_prop}
\end{align}
In the numerical calculations of Sec.~\ref{sec:numerics}, we will project the action to the 
critical trajectory $\delta_b = 0$ for all $\Lambda$.
$\vec{\phi}$ is a three-component field that should not be regarded as a fundamental 
boson but rather as a large momentum electron-hole pair that can 
scatter electrons between the hot spots of Fig.~\ref{fig:fs-hs}. The point group 
symmetry of the underlying square lattice result in an isotropic boson dispersion  
in $x$ and $y$-direction. 

This quadratic bosonic part of the action is complemented by a quartic self-interaction $u$:
\begin{align}
\Gamma_\mathrm{\phi^4}^\Lambda[\vec{\phi}]
=
u
\int \frac{ d^3 k_1}{(2\pi)^3} 
\int \frac{ d^3 k_2}{(2\pi)^3} 
\int \frac{ d^3 q}{(2\pi)^3} 
\left[ 
\vec{\phi}(k_1 + q) \cdot \vec{\phi}(k_2 - q)
\right]
\left[ 
\vec{\phi}(k_1) \cdot \vec{\phi}(k_2)
\right]\;.
\end{align}
This $\phi^4 $ term and the bosonic mass $\delta_b$ are zero in the numerical calculations of Sec.~\ref{sec:numerics} but 
will be included in the universal $\beta$-functions in Sec.~\ref{sec:flow-mass}.

%
%
%
%

In addition, electrons and spin-waves interact via a ``Yukawa-type'' Fermi-Bose vertex $g$
\begin{align}
\Gamma_\mathrm{int}^\Lambda[\bar{\psi},\psi,\vec{\phi}]
=
g \sum_{n=1}^4 \sum_{\sigma,\sigma'=\uparrow,\downarrow}
\int \frac{ d^3 k}{(2\pi)^3} 
\int \frac{ d^3 q}{(2\pi)^3} 
\left[
\vec{\phi}(q) \cdot
\bar{\psi}^{(+)}_{n,\sigma}(k+q) \vec{\tau}_{\sigma,\sigma'} 
\psi^{(-)}_{n,\sigma'}(k)
+ c.c.
\right]\;,
\label{eq:yuk}
\end{align}
which allows decay and recombination of spin-waves into electron-hole pairs.

 Finally, four-point fermion vertices are generated at the one-loop level which are believed 
 to trigger superconductivity and charge ordering, in particular. Including these 
 on top of the minimal model, will allow to detect at which scales pairing and charge ordering 
 fluctuations become important and how they modify the scaling.
 The logic here to drop these vertices for now, is to establish a complete understanding 
 of the critical state first (which so far has lacked) to then, in a subsequent step, 
 attack the more complex problem of competing instabilities and pairing retaining 
 the entire, compact Fermi surface.

We close this presentation of the model by stating that Eqs.~(\ref{eq:ferm_prop}-\ref{eq:yuk}), 
together with the regulators defined in the following,
describe a local (in time and space) quantum field theory containing only gradient operators, 
a short-range interaction that couples a metallic Fermi system to a scalar three-component boson, and a $\phi^4$ term in the bosonic field.
As we will show in the next section, singularities and non-analytic terms arise only as 
$\Lambda$ is lowered towards lower energies.

%
%

\section{Renormalization group framework}
\label{sec:RG}
We here adapt Polchinski's \cite{polchinski84} flow equation in the Legendre-transformed 
form put forward by Wetterich \cite{wetterich93} to quantum-critical metals with fluctuating 
Fermi surfaces. The main technical contribution of the present paper is the development of soft 
frequency regulators for coupled Fermi-Bose flows that (i) enable scaling towards 
flowing Fermi surfaces, that is Fermi surfaces whose shape change under the RG flow, 
and (ii) allow penetration of the universal 
quantum-critical regime at lowest energies $\Lambda\to 0$ using dynamical 
adaptation of regulators \cite{dynRG}.

A general feature of the Polchinski-Wetterich or functional RG technique \cite{metzner_review} 
are infrared regulators which are added to the quadratic parts of the 
one-particle irreducible effective action
\begin{align}
\Gamma_{R_f}^\Lambda[\bar{\psi},\psi]
&=
\sum_{n=1}^4 \sum_{m=\pm} \sum_{\sigma = \uparrow,\downarrow}
\int \! \frac{ d^3 k}{(2\pi)^3} \,
\bar{\psi}^{(m)}_{n,\sigma}(k)
\,
 R^{(m)}_{f,n} (k,\Lambda)
 \,
\psi^{(m)}_{n,\sigma}(k)\;,
\nonumber\\[2mm]
\Gamma_{R_b}^\Lambda[\vec{\phi}]
&=\frac{1}{2} 
\int \! \frac{ d^3 q}{(2\pi)^3} 
\,
R_b(\Lambda)
\,
\vec{\phi}(-q) \cdot \vec{\phi}(q)\;,
\end{align}
with the purpose to systematically suppress infrared modes of the involved quantum fields 
with energies $ < \Lambda$. At the end of the flow the regulators vanish, $R(\Lambda\to 0)
\to 0$, and the unmodified, renormalized one-particle irreducible 
action $\Gamma^{\Lambda\to 0}$ can be obtained from the flow equation 
\begin{align}
\frac{d}{d\Lambda} \Gamma^{\Lambda}[\bar{\psi},\psi,\vec{\phi}]=
\frac{1}{2}
\operatorname{Str}
\left\{
\dot{R}(\Lambda)
\left[
\Gamma^{(2)\Lambda}[\bar{\psi},\psi,\vec{\phi}]
+
R(\Lambda)
\right]^{-1}
\right\}
\;.
\label{eq:wetterich_equation}
\end{align}
with $\dot{R} = \partial_\Lambda R$.
In constructing a Polchinski-Wetterich approach to metallic quantum criticality that systematically 
decimates energy shells, a key question 
is the choice of an appropriate cutoff scheme. 
The problem defined by Eqs.~(\ref{eq:ferm_prop}-\ref{eq:yuk}) is one 
of coupled infrared singularities at different points in phase space: the electrons become 
singular at small frequencies close to the Fermi surface (at finite momenta), the 
spin waves become singular at small frequencies close to the origin of momentum space.
We will now design and implement soft frequency regulators to handle this.

A key feature of our framework is that correlations 
and fluctuations are accounted for 
\emph{systematically ordered in energy scales} and the starting point of the flow is 
always the \emph{bare action}. We believe this approach best reflects the physics 
and resolves the multiple scales 
inherent in this correlated electron problem rather well. 
Before proceeding, let us mention that the functional RG has been 
successfully applied to problems without an obvious small parameter as 
is the also case for the antiferromagnetic strange metal: 
for the 3d Ising model and O(N) models for example, extended but systematic
and quite elaborate truncations have yielded exponents comparable 
to 7-loop $\epsilon$-expansion \cite{berges_review}.

\subsection{Soft frequency regulators for flowing Fermi surfaces}
%
Frequency regulators for coupled Fermi-Bose systems at zero temperature 
exploit the fact that both, fermions and bosons become singular at the 
same point along the frequency axis, namely the origin $k_\tau = 0$ \cite{strack08}.
As we will be performing a gradient expansion to extract the flow of the 
$A$-factors in Eqs.~(\ref{eq:ferm_prop}-\ref{eq:bose_prop}), a 
further design criterion for the functional forms 
of $R^{(m)}_{f,n} (k)$ and $R_b(\Lambda)$ is the differentiability with respect to its 
(frequency and momentum) arguments. Finally, we want to allow the 
renormalized propagators, fully including the self-energy effects, to attain a scaling 
form at lowest energies. For the fermions, a functional form that achieves this is given by 
\begin{equation}
 R^{(m)}_{f,n} (k) = \left[ A^f_{\tau} i k_\tau + e^{(m)}_n(\mathbf{k},\Lambda) \right] 
 \left[ \chi^{-1} (k_\tau,\Lambda) -1 \right]
\end{equation}
 where 
\begin{equation} 
 \chi (k_\tau,\Lambda) = \frac{k_\tau^2}{k_\tau^2 + \Lambda^2} \, .
\end{equation}
 This way, the entire fermion propagator, including self-energy terms 
 \footnote{
Note that, in contrast, the $\Omega$-flow of Ref.~\onlinecite{giering-nfl} 
%
%
multiplies the \emph{bare} propagator by $\chi (k_\tau,\Lambda)$. While the mildness of this $\Omega$ regulator can be very useful in functional RG calculations with discretized frequency dependence of the fermion self-energy and the vertices, it suppresses the self-energy feedback at $ k_\tau \ll \Lambda$, which conflicts with the gradient expansion in frequency 
employed in the present paper.}, gets multiplied by a factor 
 $ \chi (k_\tau,\Lambda)$ that smoothly suppresses contributions at energies smaller than 
 the scale $\Lambda$: 
\begin{equation}
 \left(G^R_f\right)^{(m)}_n (k) = \left[ 
A^f_\tau i k_\tau + e^{(m)}_n(\mathbf{k},\Lambda) + R_{f,n}^{(m)} (k)
 \right]^{-1}
 =
 \frac{\chi (k_\tau,\Lambda)}{ A^f_{\tau} i k_\tau + e^{(m)}_n(\mathbf{k},\Lambda) } \, .
 \label{eq:reg_fermiprop}
\end{equation}
%
In the following, a superscript ${R}$ will label the regularized propagators.
Let us note in passing that the soft frequency regulators 
include excitations close to the Fermi surface in momentum space already at relatively early 
stages in the flow and thereby do not suppress Landau damping,
 as a hard momentum cutoff around the Fermi surface would 
 (see Sec.~\ref{sec:landau-damping} for a more detailed discussion).

Let us now move on to the bosonic regulator. For the bosons, a minimally invasive choice that, 
as in the fermionic regulator, identifies $\Lambda$ with frequency is 
\begin{equation}
R_b = \mathcal{A}^b_\tau \Lambda^2\;.
\end{equation}
Therewith, the infrared safe Bose propagator takes the form
\begin{equation}
 G^R_b (q) 
 = - \left[ \mathcal{A}^b_\tau q_\tau^2 
+ \mathcal{A}^b_{xy} \left( q_x^2 +   q_y^2 \right)
+ \delta_b + R_b (q) \right]^{-1}
 = \frac{-1}{\mathcal{A}^b_\tau (q_\tau^2 + \Lambda^2)
+ \mathcal{A}^b_{xy} \left( q_x^2 +   q_y^2 \right)+\delta_b}\;.
\label{eq:reg_boseprop}
\end{equation}
In this work, we follow the dynamical RG concept of Ref.~\onlinecite{dynRG}:
Apart from their explicit dependence on $\Lambda$,
the regulators also implicitly depend on the scale through running renormalization parameters.
The bosonic ``single-scale'' propagator therefore reads
\begin{equation}
 S_b (q) = \partial_\Lambda^R G^R_b (q)
 = \frac{2 \Lambda \mathcal{A}^b_\tau ( 1 - \eta^b_\tau /2)}{\left[\mathcal{A}^b_\tau (q_\tau^2 + \Lambda^2) + \mathcal{A}^b_{xy} \left( q_x^2 +   q_y^2 \right) + \delta_b \right]^2}\;,
\end{equation}
where $ \eta^b_\tau = - \partial \ln \mathcal{A}^b_\tau / \partial \ln \Lambda $ has to be fed back on the single-scale propagator
and we introduced the regulator derivative $\partial_\Lambda^R = \partial_\Lambda R\, \partial_R$.
In principle, this is true also for the fermionic single-scale propagator $ S_f = \partial^R_\Lambda G_f^R $, 
which gives rise to a coupled system of implicit equations.
In practice, however, the dynamical feedback contribution to the fermionic single-scale propagator is 
small, and vanishes in the deep infrared (see App.~\ref{sec:dyn_ferm}), 
and will therefore be dropped from the outset.


\subsection{Eight coupled flow equations}
\label{sec:big8}

We now list the flow equations for the eight flowing coupling constants 
introduced in our minimal model.
If the RG flow is pinned to criticality as in Sec.~\ref{sec:numerics}, six running couplings remain, 
as  $\delta_b = u =0 $ in that case.
The standard 
procedure is to plug in Eqs.~(\ref{eq:ferm_prop}-\ref{eq:yuk}) into 
the flow equation Eq.~(\ref{eq:wetterich_equation}) and compare 
coefficients on both sides.

The flow of the bosonic self-energy is determined by regulator derivatives of 
the ``bubble'' 
\begin{equation}
 \partial_\Lambda \Sigma_b (q) = 16 g^2 \, \partial_\Lambda^R \mathcal{S}_{C_{4v},\mathbf{q}} 
 \operatorname{Re} \left[ \int_k \left( G_{f}^{R} \right)^{(+)}_1 (k) \left( G_{f}^{R} \right)^{(-)}_1 (k-q) \right]\;,
\label{eq:bose_self}
\end{equation}
where we used a compact notation for the frequency- and momentum 
integration $\int_k f (k) = \frac{1}{(2 \pi)^3} \int \! d^3 k \, f(k)$.
$\mathcal{S}_{C_{4v},\mathbf{q}} $
denotes the symmetrization with respect to all point-group operations on $\mathbf{q}$ so that
\begin{equation*}
 \sum_{m=\pm} \sum_{n=1}^4 \int_k \left( G_{f}^{R} \right)^{(m)}_n (k) \left( G_{f}^{R} \right)^{(-m)}_n (k-q)
= 8 \mathcal{S}_{C_{4v},\mathbf{q}} \left[ \int_k \left( G_{f}^{R} \right)^{(+)}_1 (k) \left( G_{f}^{R} \right)^{(-)}_1 (k-q) \right] \, .
\end{equation*}

The fermionic self-energy flows according to
\begin{equation}
 \partial_\Lambda \Sigma_{f,n}^{(m)} (k) = - 3 g^2 \, \partial_\Lambda^R  \int_q G_b^R (q) 
 \left( G_f^R \right)_n^{(-m)} (q+k) \, .
\end{equation}
The flow of the Fermi-Bose interaction is driven by
\begin{equation}
 \partial_\Lambda g = g^3 \partial_\Lambda^R \int_q 
 \left( G_{f}^{R} \right)^{(+)}_1 (q) \left( G_{f}^{R} \right)^{(-)}_1 (q) \, G_b^R (q)\;.
 \label{eq:yuk_flow}
\end{equation}
The flow of the bosonic $\mathcal{A}^b$-factors comes from
Eq.~(\ref{eq:bose_self}) via frequency and momentum derivatives:
\begin{align}
 \partial_\Lambda \mathcal{A}^b_{\tau} &= \frac{1}{2} \left. \frac{\partial^2}{\partial q_\tau^2} \partial_\Lambda \Sigma_b (q) \right|_{q=0} \, ,\\
 \partial_\Lambda \mathcal{A}^b_{xy} &= \frac{1}{2} \left. \frac{\partial^2}{\partial q_x^2} \partial_\Lambda \Sigma_b (q) \right|_{q=0} \, .
 \label{eq:bose_flow}
\end{align}
Similarly, we have for the fermionic $A^f$-factors:
\begin{align} \label{eq:af0_flow}
 \partial_\Lambda A_{\tau}^f &= -  \left. \frac{\partial}{i \partial k_\tau} \partial_\Lambda \Sigma_{f,1}^{(+)} (k) \right|_{k=0} \, ,\\
 \partial_\Lambda A_{x}^f &= -  \left. \frac{\partial}{\partial k_x} \partial_\Lambda \Sigma_{f,1}^{(+)} (k) \right|_{k=0} \, ,\\
 \partial_\Lambda A_{y}^f &= -  \left. \frac{\partial}{\partial k_y} \partial_\Lambda \Sigma_{f,1}^{(+)} (k) \right|_{k=0} \;.
 \label{eq:fermi_flow}
\end{align}
%

 In Sec.~\ref{sec:flow-mass}, we will also take the flow of the bosonic mass $\delta_b$ and of the bosonic self-interaction $u$ into account.
 The flow equation~(\ref{eq:bose_self}) for the bosonic self-energy then acquires an additional term corresponding to a bosonic tadpole:
\begin{equation}
 \partial_\Lambda \Sigma_b (q) = 16 g^2 \, \partial_\Lambda^R \mathcal{S}_{C_{4v},\mathbf{q}} 
 \operatorname{Re} \left[ \int_k \left( G_{f}^{R} \right)^{(+)}_1 (k) \left( G_{f}^{R} \right)^{(-)}_1 (k-q) \right] - 20 u \, \partial_\Lambda^R \int_p  G_b^R (p) \, .
\label{eq:bose_self-w-mass}
\end{equation}
 This term does not affect $ \mathcal{A}^b_\tau $ and $\mathcal{A}^b_{xy} $, but counteracts the bubble contribution to the mass in
 \begin{equation} \label{eq:mass-flow}
	\partial_\Lambda \delta_b =\partial_\Lambda \Sigma_b (0) \, .
\end{equation}
The bosonic four-point vertex flows according to
\begin{equation} \label{eq:u-flow}
	\partial_\Lambda u = - 44 u^2 \partial_\Lambda^R \int_q \left[ G_b^R (q) \right]^2
	+ 4 g^4 \partial_\Lambda^R \int_k \left[ \left( G_{f}^{R} \right)^{(+)}_1 (k) \left( G_{f}^{R} \right)^{(-)}_1 (k) \right]^2 \, .
\end{equation}
The first term corresponds to a fully bosonic loop and the second to a ring of four Yukawa vertices connected by fermionic lines.
As we show in Sec.~\ref{sec:landau-damping}, the latter vanishes for $ \Lambda \ll \Lambda_\mathrm{UV} $.

Explicit expressions 
for the flow equations are in App.~\ref{sec:explicit-flow}.


\section{Numerical results: solving the flow for all $\Lambda$}
\label{sec:numerics}
In this section, we perform a comprehensive numerical analysis of the 
antiferromagnetic strange metal in $d=2$:
(i) we solve the six coupled equations (\ref{eq:yuk_flow}-\ref{eq:fermi_flow}) numerically 
covering the entire range from highest energies $\Lambda \sim \Lambda_\mathrm{UV}$ to $\Lambda\to 0$
 (For a detailed description of the numerics, see App.~\ref{sec:numerical-setup}.),
(ii) we identify three qualitatively distinct scaling regimes and crossover criteria,
(iii) we obtain scaling relations between anomalous exponents, and (iv) we show how to include 
anomalous logarithmic corrections, that lead to 
locally nested Fermi surfaces at the hot spots, into universal scaling amplitudes with 
pristine fixed-point behavior. 
Due to purely technical reasons related to numerical accuracy, we 
have run the numerics for the pinned mass case $\delta_b = 0\;\forall \Lambda$.
This shifts the numerical values of critical exponents compare to those obtained from 
Eqs.~(\ref{eq:beta-deltat}-\ref{eq:beta-eta}) but the qualitatively 
structure remains unchanged.
The actual criticality condition, correctly implemented in 
Eqs.~(\ref{eq:beta-deltat}-\ref{eq:beta-eta}) and Fig.~\ref{fig:sm-fixed}, 
is:
\begin{align}
\lim_{\Lambda \rightarrow 0} \delta_b \rightarrow 0\;.
\end{align}
$\Lambda_{\rm UV}$ is put as an upper cutoff on the $k_x$ 
integration and mimicks a curvature term that would ensures UV convergent integrals 
at initial stages of the flow. The universal infrared physics however becomes 
completely independent of $\Lambda_{\rm UV}$.
%

To characterize the flow and quantify anomalous power-law scaling, it is useful to define dimensionless 
slopes of the six flowing coupling constants. For the bosons and Yukawa vertex: 
\begin{align}
\eta^b_\tau = - \frac{\Lambda}{\mathcal{A}^b_\tau}\frac{\partial\mathcal{A}^b_\tau}{\partial \Lambda}
\;,\quad
\eta^b_{xy} = - \frac{\Lambda}{\mathcal{A}^b_{xy}}\frac{\partial\mathcal{A}^b_{xy}}{\partial \Lambda}
\;,\quad
\eta_g = - \frac{\Lambda}{g}\frac{\partial g}{\partial \Lambda}\;.
\label{eq:etas1}
\end{align}
For the electron propagator, we analogously have 
\begin{align}
\eta^f_\tau = - \frac{\Lambda}{A^f_\tau}\frac{\partial A^f_\tau}{\partial \Lambda}
\;,\quad
\eta^f_{x} = - \frac{\Lambda}{A^f_{x}}\frac{\partial A^f_{x}}{\partial \Lambda}
\;,\quad
\eta^f_{y} = - \frac{\Lambda}{A^f_{y}}\frac{\partial A^f_{y}}{\partial \Lambda}\;.
\label{eq:etas2}
\end{align}
In order to derive universal scaling functions it will be useful to define certain ratios 
of the coupling constants in which all non-universal dependencies on high-energy features 
of the theory cancel out: 
   \begin{equation}
	   \varv \equiv \frac{A^f_x}{A^f_y} \, , \quad
            \varw^2 \equiv \frac{\varv ^2}{c^2}=\frac{ \mathcal{A}^b_\tau \left(A^f_x \right)^2 }{\mathcal{A}^b_{xy} \left( A^f_\tau \right)^2 } \, , \quad
	     \tilde{g}^2 \equiv \frac{g^2}{\mathcal{A}^b_\tau \Lambda \, A^f_x A^f_y} \, ,
	     \label{eq:rescaled}
   \end{equation}
and we will use them below. The physical meaning of these variables may be stated as follows.
$\varv$ is related to the (flowing) angle between the two Fermi sheets at the hot spot 
$\alpha = \tan^{-1} \varv $, and the limit $\varv \to 0$ corresponds to two locally 
nested Fermi sheets with anti-parallel Fermi velocities (see Fig.~\ref{fig:fs-angle}). $\varw^2$ may be viewed as the ratio 
of the effective electron velocity parallel to the Fermi surface to an effective spin-wave velocity 
$c^2$. Finally, $\tilde{g}^2$ may be viewed as the ``effective coupling constant" between fermions and bosons 
relative to the canonical scaling dimension of the Yukawa vertex ($\Lambda$) and components 
of the kinetic energy of the boson ($\mathcal{A}^b_\tau$) and the fermions ($A^f_x A^f_y$).

\begin{figure}[t]
 \includegraphics[width=\linewidth]{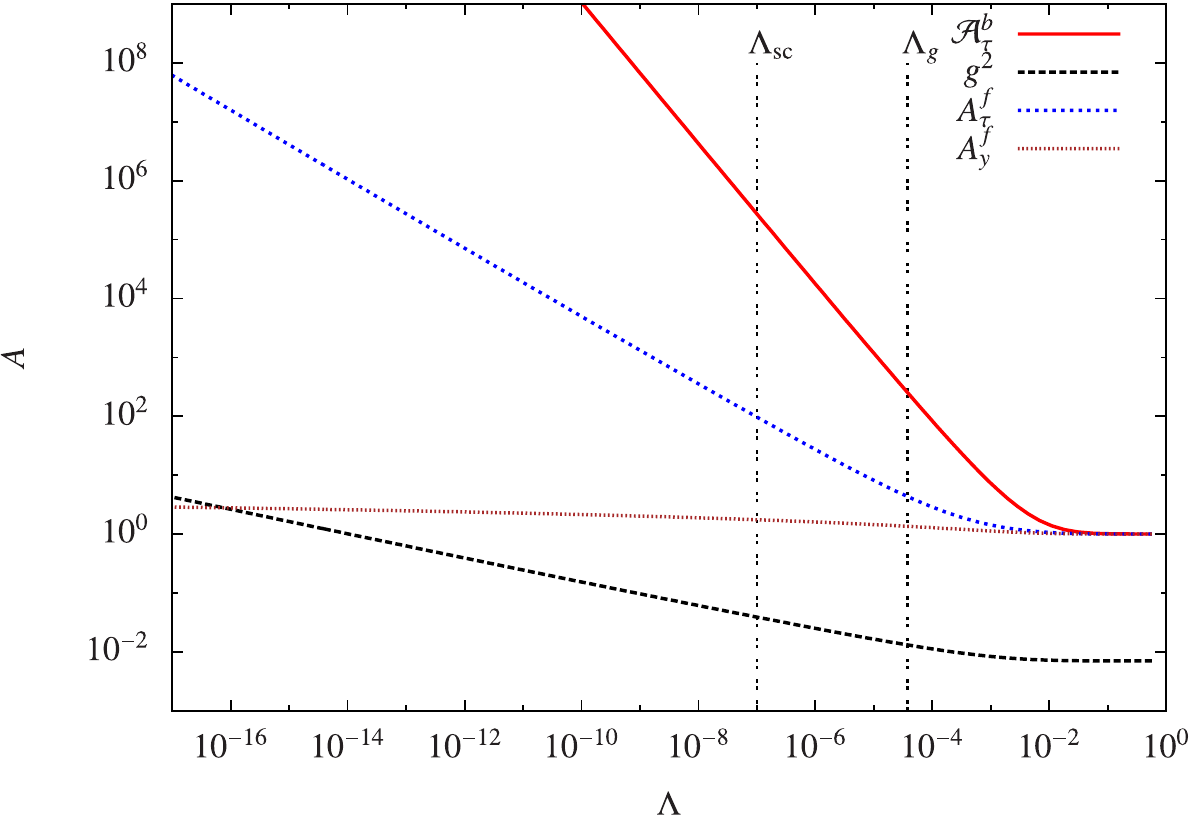}
 \caption{Flow (with pinned mass $\delta_b = 0\;\forall \Lambda$) of the running couplings with the exception of $\mathcal{A}^b_{xy}$, which is depicted in a separate plot (Fig.~\ref{fig:abxys}).
 Initial values for the flow are $\mathcal{A}^b_{\tau,\mathrm{bare}} = 1.0$, $\mathcal{A}^b_{xy,\mathrm{bare}} = 0.5$, $ g_\mathrm{bare}^2 = 0.007$, $ \varv_\mathrm{bare} =0.3$, $ A^f_{\tau,\mathrm{bare}} = A^f_{y,\mathrm{bare}} = 1.0 $, $ \Lambda_\mathrm{UV} = 0.1$. The double-dotted vertical lines mark the crossover scales between the different regimes of the RG flow explained in 
 Sec.~\ref{sec:regimes}.}
 \label{fig:as}
\end{figure}

Fig.~\ref{fig:etas} displays anomalous exponents defined in Eq.~(\ref{eq:rescaled}) from 
a numerical solution of Eqs.~(\ref{eq:yuk_flow}-\ref{eq:fermi_flow})
 indicative of power-law scaling behavior of the electrons, spin-wave, 
and the interaction vertex over several orders of magnitude in energy or cutoff 
scale $\Lambda$ converging toward universal values as $\Lambda\to 0$.
Physical implications are discussed in the key results section \ref{sec:key-results}.
The power-law flow of the corresponding couplings is depicted in Fig.~\ref{fig:as}.
\subsection{Three regimes}
\label{sec:regimes}
\begin{figure}[t]
 \includegraphics[width=\linewidth]{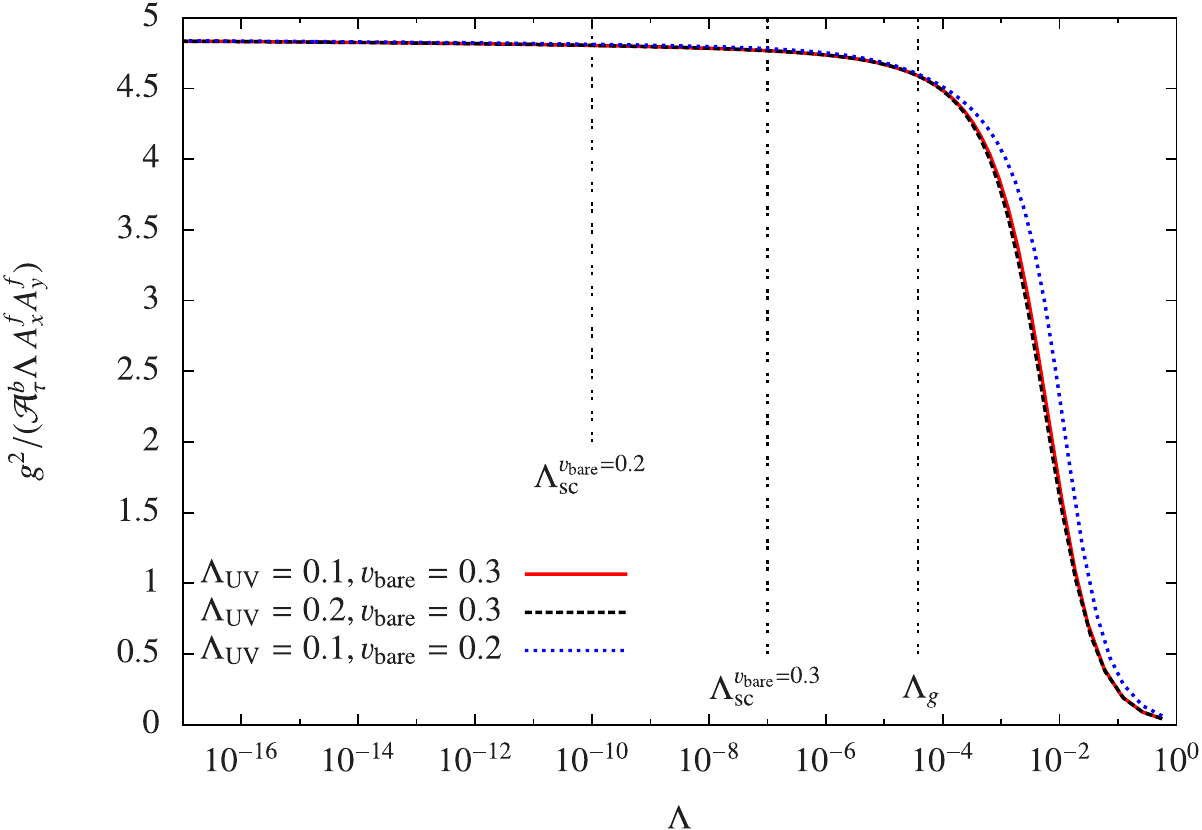}
 \caption{Flow (with pinned mass $\delta_b = 0\;\forall \Lambda$) of the coupling ratio $ \tilde{g}^2 \equiv g^2 /\left( \mathcal{A}^b_\tau \Lambda \, A^f_x A^f_y \right) $
 for $\mathcal{A}^b_{\tau,\mathrm{bare}} = 1.0$, $\mathcal{A}^b_{xy,\mathrm{bare}} = 0.5$, $g_\mathrm{bare}^2 = 0.007$, $ A^f_{\tau,\mathrm{bare}} = A^f_{y,\mathrm{bare}} = 1.0 $. The three curves correspond to different choices of $ \varv_\mathrm{bare} $ and the UV momentum
 cutoff $\Lambda_\mathrm{UV}$.
 Note that the crossover scale $ \Lambda_\mathrm{sc}$ to the anomalous logarithmic scaling depends on the angle $ \alpha_\mathrm{bare} = \tan^{-1} \varv_\mathrm{bare} $ between the bare Fermi surfaces.}
 \label{fig:g-rat}
\end{figure}
In addition to the universal exponents in the limit $\Lambda\to 0$, 
the solution of the flow equations sweeps through the entire energy range and yields 
three distinct regimes from Figs.~\ref{fig:g-rat}-\ref{fig:zs}:
\subsubsection{Non-universal regime $\Lambda_g < \Lambda < \Lambda_\mathrm{UV}$ }
\begin{figure}
 \includegraphics[width=\linewidth]{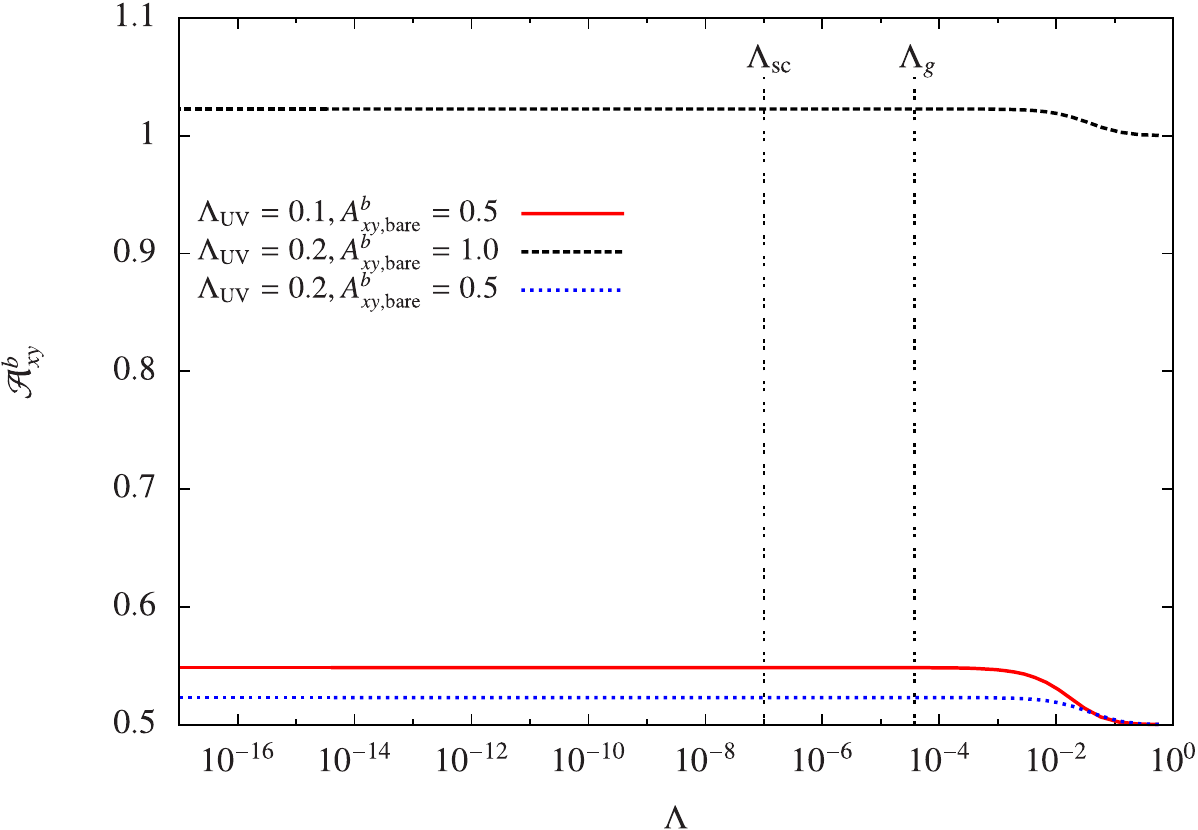}
 \caption{Flow (with pinned mass $\delta_b = 0\;\forall \Lambda$) of the bosonic momentum factor $ \mathcal{A}^b_{xy} $ for $\mathcal{A}^b_{\tau,\mathrm{bare}} = 1.0$, $ \varv_\mathrm{bare} =0.3$, $ g_\mathrm{bare}^2 = 0.007$, $ A^f_{\tau,\mathrm{bare}} = A^f_{y,\mathrm{bare}} = 1.0 $. The three curves correspond to different choices of $ \mathcal{A}^b_{xy,\mathrm{bare}} $ and the UV cutoff $\Lambda_\mathrm{UV}$.}
 \label{fig:abxys}
\end{figure}
The flow is quiet for $\Lambda>\Lambda_\mathrm{UV}$ as all excitations, electrons and bosons, are pushed off-shell through large values 
of the cutoff scale $\Lambda$ and the kernels formed by regularized 
Green's functions in the six equations Eqs.~(\ref{eq:yuk_flow}-\ref{eq:fermi_flow}) are 
small.

At $ \Lambda \sim \Lambda_\mathrm{UV} $, we enter a regime where the anomalous dimensions start to grow (see Fig.~\ref{fig:etas}), i.e.\ where renormalizations set in.
Starting from sub-power-law behavior, the anomalous dimension $\eta^b_\tau$ associated with $ \mathcal{A}^b_\tau$ increases rapidly and eventually exceeds the
Hertz-Millis Landau damping value $ \eta^b_\tau = 1$ obtained by integrating out bare fermions (see Sec.~\ref{sec:landau-damping}).
At lower scales, the flow preserves this super-Landau-damping characteristics.
All other running couplings have anomalous dimensions (much) smaller than $1$.

Above the scale $ \Lambda_g $, the precise values of the running couplings and the associated anomalous dimensions depend on the bare couplings.
This is clearly visible in the flow of $\mathcal{A}^b_{xy} $, which is depicted in Fig.~\ref{fig:abxys}.
This quantity only is renormalized at scales $ \Lambda_g < \Lambda < \Lambda_\mathrm{UV}$.
Its growth at these scales is a direct consequence of the finiteness of the UV momentum cutoff $\Lambda_\mathrm{UV}$
or curvature terms in the electron dispersion.
Namely, in the absence of such a UV scale, the residue theorem prohibits any renormalization of $ \mathcal{A}^b_{xy}$ (cf.~Sec.~\ref{sec:landau-damping}).

We conclude that the flow is non-universal in the regime $ \Lambda_g < \Lambda < \Lambda_\mathrm{UV} $. 
In particular, the different values of the UV momentum cutoff $\Lambda_\mathrm{UV}$ and the initial 
value of the Fermi surface angle $\varv_\mathrm{bare}$ result in different values 
of the running couplings in this regime  
(cf.\ Figs.~\ref{fig:g-rat} and \ref{fig:abxys}).
   
 \subsubsection{Crossover regime $\Lambda_\mathrm{sc} < \Lambda < \Lambda_g$}
\begin{figure}[t]
 \includegraphics[width=\linewidth]{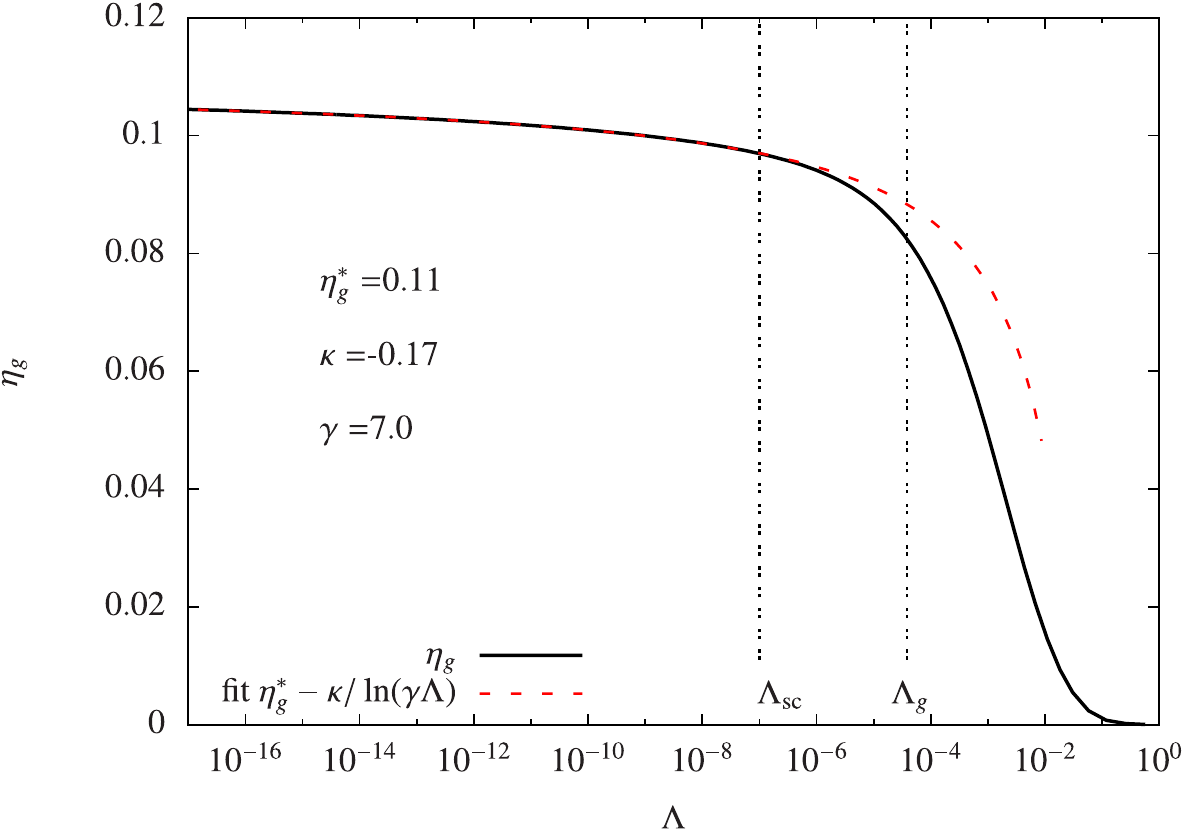}
 \caption{Flow (with pinned mass $\delta_b = 0\;\forall \Lambda$) of the anomalous exponent of the Yukawa vertex (solid line) $\eta_g $ for $\mathcal{A}^b_{\tau,\mathrm{bare}} = 1.0$, $\mathcal{A}^b_{xy,\mathrm{bare}} = 0.5$, $ g_\mathrm{bare}^2 = 0.007$, $ \varv_\mathrm{bare} =0.3$, $ A^f_{\tau,\mathrm{bare}} = A^f_{y,\mathrm{bare}} = 1.0 $, $ \Lambda_\mathrm{UV} = 0.1$ and fit (dashed line) to Eq.~(\ref{eq:def-pssc}). Note that especially the (non-universal) fit parameter $\gamma$ is difficult to determine, and that therefore also $\eta_g^\ast$ may be slightly overestimated in the fit above.}
 \label{fig:pseudoscale}
\end{figure}
At $ \Lambda < \Lambda_g $, the flow of the rescaled Yukawa coupling $\tilde{g}$ looses its memory of the bare Fermi surface angle and the UV cutoff $\Lambda_\mathrm{UV}$:
In Fig.~\ref{fig:g-rat}, the curves of $\tilde{g}$ corresponding to different values of $ \varv_\mathrm{bare}$ and $\Lambda_\mathrm{UV}$ coalesce.
However, the flow of $\tilde{g}$ only becomes fully universal at a later stage of the flow, since it still strongly depends on $g_\mathrm{bare}$ directly below $\Lambda_g$. A criterion for the end of this 
regime is the intersection of the anomalous exponent of the bare vertex $\eta_g$, 
black, solid line in Fig.~\ref{fig:pseudoscale} with its universal fitting function (red, dashed) at 
$\Lambda \approx 10^{-7}$ for $ \varv_\mathrm{bare} = 0.3 $ (we describe this in more detail below).
 \begin{figure}
 \includegraphics[width=\linewidth]{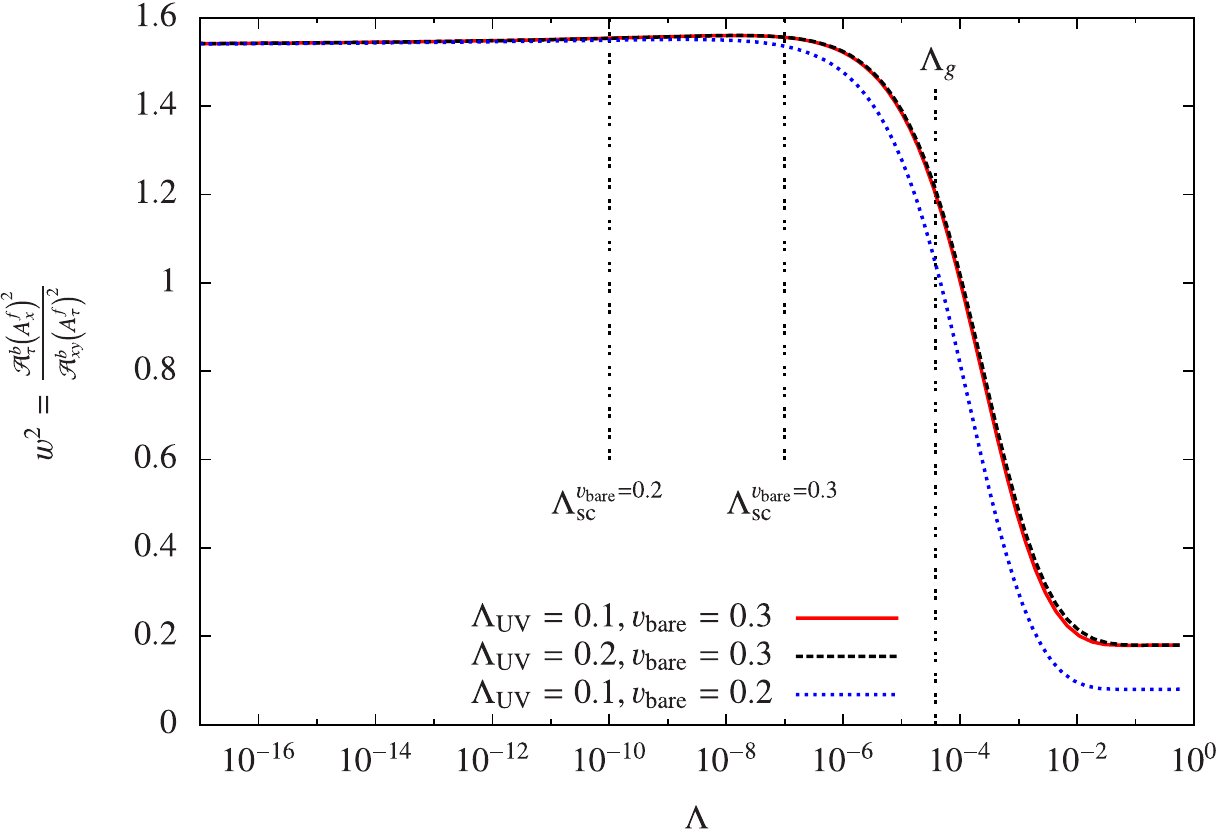}
 \caption{Flow (with pinned mass $\delta_b = 0\;\forall \Lambda$) of the velocity ratio $ \varw^2 $ for $\mathcal{A}^b_{\tau,\mathrm{bare}} = 1.0$, $ \mathcal{A}^b_{xy} =0.5$, $ g_\mathrm{bare}^2 = 0.007$, $ A^f_{\tau,\mathrm{bare}} = A^f_{y,\mathrm{bare}} = 1.0 $. The three curves correspond to different choices of $ \varv_\mathrm{bare} $ and the UV momentum cutoff $\Lambda_\mathrm{UV}$.
 For further explanation, see the text.
 Note that the crossover scale $ \Lambda_\mathrm{sc}$, at which $\varw^2$ has relaxed to its universal value, depends on the angle $ \alpha_\mathrm{bare} = \tan^{-1} \varv_\mathrm{bare} $ between the bare Fermi surfaces.}
 \label{fig:w-rat}
\end{figure}
A second criterion yielding the same crossover scale to the fully universal regime is 
the scale when the effective velocity ratio $ \varw^2$ has saturated to its 
universal value of about $5/4$
for different initial UV-cutoffs. 
For the black-dashed line and red-solid line in Fig.~\ref{fig:w-rat}, this yields the same crossover scale scale $\Lambda_\mathrm{sc}$.
The individual couplings  $ \mathcal{A}^b_{xy} $ and $ \mathcal{A}^b_\tau / ( A^f_\tau A^f_y )^2 $ that make up $\varw^2$
 remain \emph{non-universal} constants.
 It is only in the ratio $\varw^2$ that they cancel exactly.
 Let us note here that the product $ A^f_x A^f_y $ is a non-universal constant at all scales and does not 
renormalize at all,
 \begin{equation} \label{eq:screl-fxy}
 \partial_\Lambda \left( A^f_x A^f_y \right) = 0
 \;,\;\;\; \mathrm{and }\;\;\;\;
  \eta^f_x = - \eta^f_y\;.
 \end{equation}
\begin{figure}
 \includegraphics[width=\linewidth]{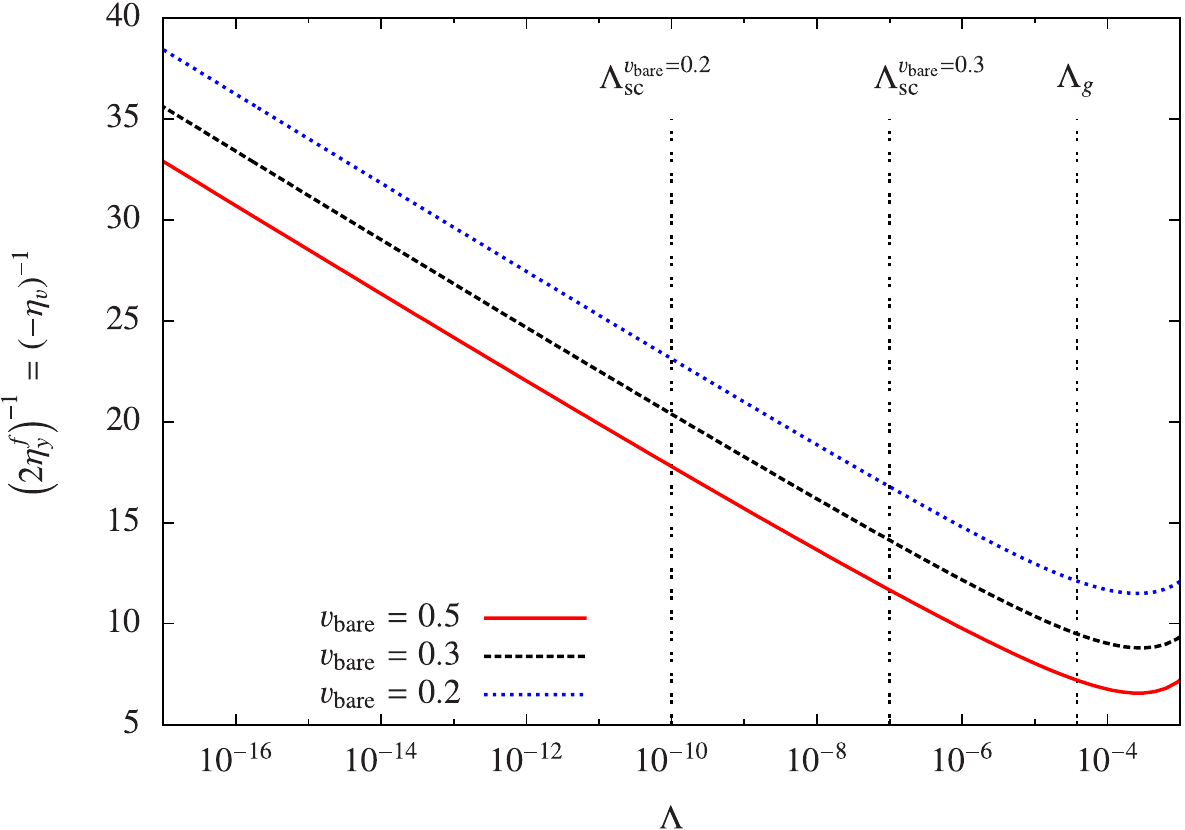}
 \caption{Inverse of the (vanishing) anomalous dimension (flow for pinned mass $\delta_b = 0\;\forall \Lambda$) associated with the Fermi surface angle $\varv$ for chosen parameters $\mathcal{A}^b_{\tau,\mathrm{bare}} = 1.0$, $\mathcal{A}^b_{xy,\mathrm{bare}} = 0.5$, $ g_\mathrm{bare}^2 = 0.007$, $ A^f_{\tau,\mathrm{bare}} = A^f_{y,\mathrm{bare}} = 1.0 $, $ \Lambda_\mathrm{UV} = 0.1$. 
 The curves are straight lines in agreement with the ansatz
 $\eta = - \frac{\kappa}{  \ln (\gamma \Lambda)} $ [cf.\ Eq.~(\ref{eq:def-pssc})].
 Their slopes correspond to $\kappa$, and their offsets to $ \kappa \ln \gamma$.
 According to our data, $\kappa$ is universal, while $ \gamma $ is highly non-universal.}
 \label{fig:eta-v}
\end{figure}
We emphasize that although these (mild) memory effects disappear 
completely only at $\Lambda_\mathrm{sc}$ and below, they cancel already earlier 
in the flow (at $\Lambda_g$)  in the ratio $\tilde{g}^2$ 
of Eq.~(\ref{eq:rescaled}) and plotted in Fig.~\ref{fig:g-rat}. 
From Fig.~\ref{fig:w-rat}, we infer that the remaining non-universalities depend on $\varv_\mathrm{bare}$.
This parameter plays a more important role for the renormalization of the fermionic renormalization constants that enter in $\varw$ than for the bosonic ones and the Yukawa vertex.
Since the renormalizations of the fermionic couplings cancel in $\tilde{g}$, but not in $\varw$,
the curves for $\varw$ in Fig.~\ref{fig:w-rat} coalesce at a lower scale than those for $\tilde{g}$ in Fig.~\ref{fig:g-rat}.


\subsubsection{Universal regime $0 < \Lambda < \Lambda_\mathrm{sc}$}
 %
 %
 At scales below $ \Lambda_\mathrm{sc} $, both $ \varw^2 $ and $ \tilde{g}^2$ have saturated to their IR values, 
 while $\varv$ flows logarithmically to zero.
 Also that last property is universal, see Figs.~\ref{fig:flow-v} and \ref{fig:eta-v}.
In other words, we find a pristine fixed point in the variables $ \varw $ and $ \tilde{g}$, that {\it requires a finite, but 
logarithmically decaying $ \varv $}.
\begin{figure}
 \includegraphics[width=\linewidth]{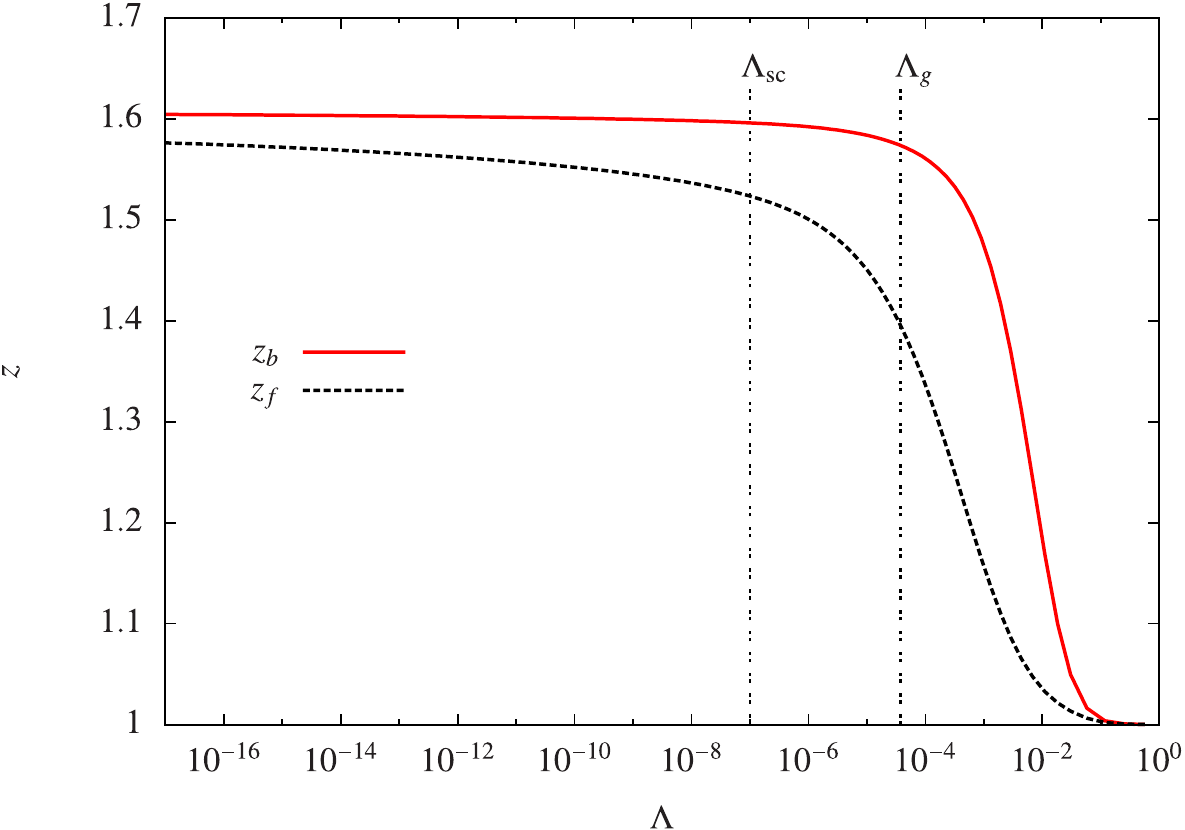}
 \caption{Flow (with pinned mass $\delta_b = 0\;\forall \Lambda$)  of dynamical exponent for bosons ($ z_b $) and fermions ($z_f$) 
 according to the additive definition $z_f = 1 + \eta^{f}_\tau - \eta^{f}_y$ 
 and $z_b = 1 + \frac{\eta^{b}_\tau - \eta^{b}_{xy}}{2}$ below
  Eq.~(\ref{eq:dynamical}). Plotted for: 
 $\mathcal{A}^b_{\tau,\mathrm{bare}} = 1.0$, $ \varv_\mathrm{bare} =0.3$, 
 $g_\mathrm{bare}^2 = 0.007$, $\mathcal{A}^b_{xy} = 0.5$,
 $ \Lambda_\mathrm{UV} =0.1$, $ A^f_{\tau,\mathrm{bare}} = A^f_{y,\mathrm{bare}} = 1.0 $. The three curves correspond to different choices of $ \mathcal{A}^b_{xy,\mathrm{bare}} $ and $\Lambda_\mathrm{UV}$.}
 \label{fig:zs}
\end{figure}
Let us now look at the scaling at this fixed point. 
In $ \eta^b_\tau $, logarithmic contributions are small and one has $ \eta^{b \ast}_\tau \approx 1.22 $ 
in the universal regime.
This gives rise to the values for the dynamical exponent $ z_b = 1.61 $ (for pinned mass flows). 
In the key results section \ref{sec:key-results}, we quote the slightly reduced values from flows 
with bosonic mass $z_b^\ast = 1.53$.
After $\varw^2$ has saturated, the flow of the non-constant running couplings is of mixed power-law logarithmic type:
 The corresponding anomalous dimensions behave as
\begin{equation} \label{eq:def-pssc}
	\eta = \eta^\ast - \frac{\kappa}{  \ln (\gamma \Lambda)}  \, .
\end{equation}
If $\eta^\ast=0$, the anomalous logarithmic contributions in Eq.~(\ref{eq:def-pssc}) are relevant and play a crucial role in the deep infrared limit. In all relevant anomalous logarithmic terms, we find $ \kappa $ to be universal, whereas $ \gamma $ takes on non-universal values (see Fig.~\ref{fig:eta-v}).

Our scaling analysis in Sec.~\ref{sec:orig-screl} reveals that relevant logarithmic contributions originate from the flow of the intersection angle between the two Fermi surfaces as was already pointed out by Abanov and Chubukov \cite{abanov00}.
However, here we have argued
that the logarithms are consistent with 
the pristine fixed-point of $\tilde{g}$ and the velocity ratio 
$\varw^2$ rather than leading to a breakdown of scaling as was found by Metlitski and Sachdev 
within the field-theoretical renormalization group \cite{MMSS10b}.

\subsection{Scaling relations and universal properties}
\label{sec:universal}

The rescaled Yukawa coupling ratio
 $ \tilde{g}^2 \equiv g^2 /\left( \mathcal{A}^b_\tau \Lambda \, A^f_x A^f_y \right) $
 takes on the constant, universal value $ 4.9 $ (see Fig.~\ref{fig:g-rat}).
 Together with Eq.~(\ref{eq:screl-fxy}) this further 
 implies the scaling relation Eq.~(\ref{eq:screl-g-abt}).
This relation is valid up to logarithmic corrections and will be derived analytically in Sec.~\ref{sec:orig-screl}.
Probably due to the smallness of the corresponding anomalous dimension $ \eta_g^\ast \approx 0.11 $, however, 
logarithmic corrections are more clearly visible in $\eta_g$ 
(see Fig.~\ref{fig:pseudoscale}). These logarithmic contributions here are 
 subdominant, since the power-law scaling (non-zero $ \eta_g^\ast $) is the 
 stronger singularity.

Let us now turn to the fermionic sector. Here, our first observation is that 
$\varv \equiv A^f_x / A^f_y $ logarithmically flows to zero (see Fig.~\ref{fig:flow-v}).
 In contrast to the logarithmic corrections in $ \eta_g$, these logarithms are relevant.
The infrared behavior of the fermionic frequency term is of mixed ``power-law+logarithmic'' type, 
as can be seen from the curve for $ \eta^f_\tau $ in Fig.~\ref{fig:etas}. 
In the deep IR, this anomalous dimension saturates to $ \eta^{f \ast}_\tau = \eta^b_\tau /2 $.
In the flow of $ A^f_\tau $, logarithmic corrections to the power law now play a crucial role: 
Since $ \varw $ flows to a constant and since $ \eta^f_x + \eta^f_y = \eta^b_{xy} =0$ in the universal regime,
we get additionally the identity Eq.~(\ref{eq:fermbos-screl}).
 Again, we will refer to Sec.~\ref{sec:orig-screl} for an analytical derivation of this scaling relation. 
 Since corrections to the power law are found to be negligible in $ \mathcal{A}^b_\tau $, logarithmic contributions 
 on the right-hand side of Eq.~(\ref{eq:fermbos-screl}) have to cancel.
 In particular, the leading logarithm in $ \eta^f_y $  must be canceled by the logarithmic contributions in $ \eta^f_\tau $.
 Although this logarithmic term appears as a correction to the power law for $A^f_\tau$, it has played 
 a crucial role for the deep IR properties.

We now list the key properties of the universal regime (values 
given for pinned bosonic mass $\tilde{\delta}=0$):
(i) In the bosonic sector, $ \mathcal{A}^b_{xy} $ has saturated in the anomalous scaling regime and $ \eta^{b \ast}_\tau \approx 1.22 $ is universal.
	  This gives rise to a universal bosonic
 dynamical exponent $ z^\ast_b \approx 1.61 $.
(ii) By virtue of the scaling relation~(\ref{eq:screl-g-abt}), the Yukawa coupling obeys a universal power law with the exponent $ \eta^\ast_g \approx 0.11 $ at low scales.
(iii) In the fermionic sector, $ \varv \equiv A^f_x / A^f_y $ flows to zero as $ 1 / \ln (\gamma \Lambda) $. The dynamical exponent $ z_f $ logarithmically approaches $ z^\ast_b $
	  (see Fig.~\ref{fig:zs}).
 In the deep IR, $ z_b =z_f$ is protected by the scaling relation Eq.~(\ref{eq:fermbos-screl}).
(iv) The velocity ratio $\varw^2$ defined in Eq.~(\ref{eq:rescaled})
 takes on a constant, universal value $ \varw^\ast = 1.24 \approx 5/4 $ at lowest scales, 
 similar to Ref.~\onlinecite{sur14}, wherein $ \varw^\ast = 2/3 $. 
(v) Also the rescaled Yukawa coupling $ \tilde{g}$ flows to a universal constant $  \tilde{g}^\ast = 2.2 \approx 11/5 $.

\subsection{Accessibility of the fixed-point at weak coupling}
\label{subsec:weak}
%
	 Our fixed point is reached for small initial, nonzero bare couplings $ g_\mathrm{bare} $ at the expense 
	 of only reaching it very late in the flow, when only a reduced amount of phase space available.
%
	 In practice, we have to ensure that our results are not plagued by strong-coupling effects already at high scales,
	 i.e.\ that the kinetic energies $ E^f_\mathrm{kin} $ and $ E^b_\mathrm{kin} $ of the bosons exceeds the typical energy scale $E_\mathrm{int}$ of the interaction.
	 Although the fermionic dispersion $e^{(m)}_n (\mathbf{k}) $ is not bounded from above in our model,
	 the scale $\Lambda$ provides a such an upper bound, since the predominant contributions of fermionic lines in the diagrams are at energies $ e^{(m)}_n \sim \Lambda $.
	 Similarly, the predominant contributions from bosonic lines correspond to $ E^b_\mathrm{kin} \sim \Lambda^2 $.

	 If we now require that $ E^f_\mathrm{kin} , E^b_\mathrm{kin} > E_\mathrm{int}$ at scales $ \Lambda \geq \Lambda_\mathrm{UV}$ larger than the UV cutoff,
	 $ g_\mathrm{bare} \leq 0.01 $ for the UV cutoffs chosen in our numerics. As we have chosen $ g_\mathrm{bare}^2 = 0.007$ in our numerical calculations, we observe that we can delay 
	 renormalizations of the Yukawa interaction to small $\Lambda$.
         We emphasize that this argument works only if the  angle between the bare Fermi surfaces is not too small.
	 For $ \varv_\mathrm{bare} \geq 0.2 $, we find that  the rescaled Yukawa coupling $ \tilde{g}$ monotonically grows and eventually saturates to its universal value
	 $ \tilde{g}^\ast$ (see Fig.~\ref{fig:g-rat}).
	 At very low values $ \varv_\mathrm{bare}$, i.e.\ in a situation where the two bare Fermi surfaces are already almost antiparallel when beginning the flow,
	 $\tilde{g}$ would experience a pronounced strong-coupling peak in the non-universal regime prior to saturating.
         From the absence of such peaks in Fig.~\ref{fig:g-rat}, we conclude that the 
         fixed point is approached from relatively weak coupling without
	 crossing any strong-coupling regime for $ g_\mathrm{bare} = 0.007$.
	 Together with the arguments by Salmhofer \cite{dynRG}, who showed that 
	 the 1-PI scheme of the functional RG (used here) with 
	 $\Lambda$-dependent adaptation performs well even at moderate coupling, we take 
	 this is as encouraging evidence that our results are qualitatively robust.
	 
\section{Analytical results: extracting universal scaling functions}
\label{sec:orig-screl}

In this section, we underpin our numerical data of the previous section with 
analytical considerations from the six coupled flow equations (\ref{eq:yuk_flow}-\ref{eq:fermi_flow}) in Sec.~\ref{sec:big8}.
We show that the key qualitative features detected in the numerics, 
such as scaling relations and universal values of exponents and coupling ratios,
can be derived from explicit analytical expressions of the flow equations.
We will retain a rescaled bosonic mass term 
$\tilde{\delta} = \delta_b /\mathcal{A}^{b}_\tau \Lambda^2 $ in the bosonic propagators with a view toward
Sec.~\ref{sec:flow-mass}.
It is convenient to use a rescaled coordinate system
 \begin{align}
& y_\tau = k_\tau / \Lambda \, , \quad y_x = A^f_x k_x / (A^f_\tau \Lambda) \, , \quad y_y = A^f_y k_y / (A^f_\tau \Lambda) \, 
	 \nonumber\\
	\int &\frac{ d k_\tau}{2\pi} \int \frac{ d k_x}{2\pi}\int \frac{ d k_y}{2\pi}
\to 
\frac{\Lambda^3 \left(A^f_\tau\right)^2}{A^f_x A^f_y}
\int \frac{ d y_\tau}{2\pi} \int \frac{ d y_x}{2\pi}\int \frac{ d y_y}{2\pi}
	 \label{eq:y_coordinates}
 \end{align}
and write the regulated electron and Bose propagators 
Eqs.~(\ref{eq:reg_fermiprop},\ref{eq:reg_boseprop}) as
 \begin{align}
	 \left( G_f^R \right)^{(\pm)}_1&=  \frac{1}{\Lambda A^f_\tau} \frac{\chi (y_\tau,1)}{\left[ i y_\tau + y_x \pm y_y \right]}
	 \nonumber\\
	 G^R_b& = \frac{1}{ \Lambda^2 \mathcal{A}^b_\tau} \frac{-1}{\left[ y_\tau^2 + \varw^{-2}
	  \left( y_x^2 + \varv^2 \, y_y^2 \right) + 1 + \tilde{\delta} \right]} \, ,
	  \label{eq:y_props}
 \end{align}
with $\varv$ and $\varw^2$ defined in Eq.~(\ref{eq:rescaled}).
 %
 The scale-dependent coordinates Eq.~(\ref{eq:y_coordinates}) remove the logarithmically vanishing 
 angle between the Fermi surfaces $\varv$ from the fermionic propagators and 
 transfers it from the fermionic to the bosonic lines. The limit $ \varv \to 0 $ amounts to 
 neglecting the
 $ y_y^2 $ term on the bosonic lines. This is only legal in the ultimate deep infrared limit $\Lambda\to 0$; the data in Fig. \ref{fig:flow-v} shows that $\varv$ vanishes logarithmically slowly and even for $\Lambda \sim 10^{-16}$ it is still sizable $\varv \sim 0.03$. 
 
 It is dangerous to associate the limit $\varv\to 0$ 
 in Eq.~(\ref{eq:y_props}) with a ``one-dimensional boson'' that disperses in the $x$-direction only.
 Rather, the $\Lambda$-dependence of $\varv$ reflects a strong $k_x$ dependence of $\varv$ such that 
 considering a limit $\varv\to 0$ without also taking $y_x \to 0$ is generally illegal. 
 Even in the cosmetic choice of coordinate system Eq.~(\ref{eq:y_coordinates}) the boson (and the fermion) 
 should be regarded as dispersing fully in two directions of space. The physically observable 
spin susceptibility given in Sec.~\ref{sec:key-results} is manifestly two-dimensional and fulfills the point-group symmetry, that is, disperses 
 homogeneously in $q_x$ and $q_y$ direction. The physical electron also 
 disperses in both directions in (momentum) space complemented by small logarithmic corrections.
 
 \subsection{Mixed fermion-boson loops: Yukawa vertex, fermion self-energy}
 \label{sec:g-abt-scr}
 
In a prototypical diagram with fermions and bosons evaluated 
 in the coordinate system Eq.~(\ref{eq:y_coordinates},\ref{eq:y_props}), the prefactors of the propagators
 and the integration measure can be absorbed into the rescaled Yukawa vertex $\tilde{g}$, 
 which attains a universal fixed-point value $\tilde{g}^\ast$ (see Fig.~\ref{fig:g-rat}).
 The integration kernels now only depend on (i) the coupling ratio $\varw$ (which attains a universal fixed-point value $\varw^\ast$ as $\Lambda \to 0$ per Fig.~\ref{fig:w-rat}), and (ii) on $\eta^b_\tau$ through the bosonic single-scale propagator (which also attains a universal fixed-point value 
$\eta^{b\ast}_\tau$ as $\Lambda \to 0$ per Fig.~\ref{fig:etas}). 
 %
A case in point is the anomalous exponent of the Yukawa vertex 
  \begin{equation} \label{eq:eta-g-ana}
	 \eta_g =  \tilde{g}^2 T_g (\varv, \varw,\eta^b_\tau, \tilde{\delta})
 \end{equation}
 where the dimensionless scaling function
 \begin{align} 
 \label{eq:tg} \notag
T_g (\varv, \varw,\eta^b_\tau, \tilde{\delta}) &= - \frac{1}{8 \pi^3} \int_{-\infty}^{\infty}d y_\tau  \int_{-\infty}^{\infty} d y_x  
	 \int_{-\infty}^{\infty} d y_y \,\, 
	 \frac{1}{ i y_\tau + y_x + y_y} \, \frac{1}{ i y_\tau + y_x - y_y}
	 \\ \notag
	 & \quad \times 
	 \left\{  \frac{4 y_\tau^4}{\left(y_\tau^2 +1 \right)^3} \, \frac{1}{y_\tau^2 + \varw^{-2}
	 \left( y_x^2 + \varv^2 \, y_y^2 \right) + 1 + \tilde{\delta} }
	 +\frac{y_\tau^4}{\left( y_\tau^2 +1 \right)^2 } \, 
	 \frac{2 - \eta^b_\tau}{\left[ y_\tau^2 + \varw^{-2}
	 \left( y_x^2 + \varv^2 \, y_y^2 \right) + 1 + \tilde{\delta} \right]^2} \right\} \, ,\\ \notag
	 & = \frac{\varw}{8 \pi} \int_{- \infty}^{+ \infty} \! d y_\tau \,
	 \left[ \frac{1}{\left| y_\tau \right| \sqrt{1 + y_\tau^2 + \tilde{\delta}} + \varw \left( 1 + y_\tau^2 + \tilde{\delta} \right)} \;
	 \frac{4 y_\tau^4}{\left( y_\tau^2 + 1 \right)^3}
	  \right. \\
	 & \qquad + \left. 
	    \frac{ \left| y_\tau \right| + 2 \varw \sqrt{1 + y_\tau^2 + \tilde{\delta}}}{
		 \left( 1 + y_\tau^2 + \tilde{\delta} \right)^{3/2}
	  \left( \left| y_\tau \right| + \varw \sqrt{ 1 + y_\tau^2 + \tilde{\delta} } \right)^2 } \;
	 \frac{y_\tau^4}{\left( y_\tau^2 + 1 \right)^2} \, \left( 1 - \frac{\eta^b_\tau}{2} \right) \right]
 \end{align}
 yields the universal number $T_g(0,\varw ^\ast, \eta^{b,\ast}_\tau, 0) = 0.027$ in the deep infrared,
 when the remaining frequency integral is calculated numerically.
  Together with $\tilde{g}^\ast = 2.2 $ this  yields the asymptotic value 
  $\eta_g^\ast = 0.11$, toward which $\eta_g$ is 
  converging in Fig.~\ref{fig:etas} as $\Lambda \to 0$.
 Since the bosonic dynamical exponent $ z_b^\ast $ is universal,
  $ \eta^{b\ast}_\tau $ and -- by virtue of the scaling relation~(\ref{eq:screl-g-abt}) -- also $ \eta_g^\ast$ should be universal.
  As a consequence, $ \varw^\ast$ must be finite and universal in order to allow for a universal $z^\ast_b$, 
  which we confirmed to high accuracy in Fig.~\ref{fig:w-rat}.
Moreover, we also find $\varw \approx \varw^\ast$ and $ \eta^b_\tau \approx \eta^{b\ast}_\tau $ to high accuracy 
throughout the regime of anomalous logarithmic scaling. This implies that the (irrelevant) logarithmic corrections in $\eta_g$ are induced by corrections to Eq.~(\ref{eq:tg}) for finite $\varv$.

  The constancy of the coupling ratio $\varw$ defined in Eq.~(\ref{eq:rescaled}) also directly gives rise to the second scaling relation Eq.~(\ref{eq:fermbos-screl}) as follows:
  Since $ A^f_x A^f_y $ and $\mathcal{A}^b_{xy} $ are non-universal constants,
  the ratio $ \mathcal{A}^b_\tau / \left( A^f_\tau A^f_y \right)^2 $ must be a non-universal constant as well, 
  which can be rephrased as $\eta^b_\tau /2 =  \eta^f_\tau + \eta^f_y $.
  
  We continue with the proof of the absence of power-law scaling in the fermionic momentum dispersion.
  To this end, we apply our change of coordinates and rescaling to Eq.~(\ref{eq:fermi_flow}) and obtain
  \begin{equation} \label{eq:eta-f-y-ana}
	 \eta^f_y = 3 \tilde{g}^2 T^f_y (\varv, \varw,\eta^b_\tau,\tilde{\delta}) \, .
 \end{equation}
 We now show that this scaling function vanishes 
 in the deep IR, that is, $T^f_y(\varv \to 0, \varw^\ast, \eta^{b\ast}_\tau,0) \to 0$.
 The explicit expression reads
 \begin{align} 
 	 \label{eq:tfy} \notag
	 T^f_y (\varv, \varw,\eta^b_\tau,\tilde{\delta}) &= -  \int_y
	 \frac{1}{\left( i y_\tau + y_x - y_y \right)^2}  \\ \notag
	 & \quad \times 
	 \left\{  \frac{2 y_\tau^2}{\left(y_\tau^2 +1 \right)^2} \, \frac{1}{y_\tau^2 + \varw^{-2}
	 \left( y_x^2 + \varv^2 \, y_y^2 \right) + 1 + \tilde{\delta} } \right.\\
	 & \qquad \left. +\frac{y_\tau^2}{y_\tau^2 +1 } \, \frac{2 - \eta^b_\tau}{\left[ y_\tau^2 + \varw^{-2}
	 \left( y_x^2 + \varv^2 \, y_y^2 \right) + 1 + \tilde{\delta}  \right]^2} \right\} \, .
 \end{align}
  In the curly brackets on the right-hand side of this equation, the first and second term correspond to the single-scale propagator being a fermionic or bosonic line, respectively.

  This integral still converges, i.e.\ taking the limit $ \varv \to 0 $ prior to calculating the loop integral does not result in a divergence.
  Since $\tilde{g}^2$ flows to a universal constant according to the last subsection~\ref{sec:g-abt-scr}
  and since the threshold function $T^f_y$ neither explicitly nor implicitly depends on the scale,
  we will only be able to recover the power-law contribution $ \eta^{f\ast}_y $ from Eq.~(\ref{eq:eta-f-y-ana}).

  In Eq.~(\ref{eq:tfy}), Fubini's theorem applies also in the limit $\Lambda_\mathrm{UV} \to \infty $, and the $y_y$ integral may be performed first.
  But then we observe that the integrand has only poles in one half of the complex $y_y$ plane and that therefore $ T^f_y $ vanishes.
  As a consequence, also $\eta^f_y$ vanishes since $ \tilde{g}^\ast $ is finite.
  Hence we conclude that the IR behavior of the fermionic dispersion must be governed by logarithmic scaling of the same type as in the second term $ - \kappa / \ln ( \gamma \Lambda) $
  in Eq.~(\ref{eq:def-pssc}).

  If derivatives of the regulator with respect to the external frequency  were neglected in $ \dot{A}^f_\tau $,
  also $ \eta^f_\tau $ would vanish in the universal regime.
  This can easily be seen from taking the deep IR limit in
  \begin{equation} \label{eq:eta-f-0-ana}
	 \eta^f_\tau = 3 \tilde{g}^2 T^f_\tau (\varv, \varw,\eta^b_\tau,\tilde{\delta}) \, , 
 \end{equation}
 where $  T^f_\tau = \hat{T}^f_\tau - T^f_y $
 with the regulator-induced part
 \begin{align}  \notag
	 \hat{T}^f_\tau (\varv, \varw,\eta^b_\tau,\tilde{\delta}) &= \int_y
	 \frac{1}{ i y_\tau + y_x - y_y } \\ \notag
	 & \quad \times 
	 \left\{  \frac{4 i y_\tau}{\left(y_\tau^2 +1 \right)^2} \left( 1 - 2 \frac{y_\tau^2}{y_\tau^2 + 1 } \right)
		\, \frac{1}{y_\tau^2 + \varw^{-2}
		\left( y_x^2 + \varv^2 \, y_y^2 \right) + 1 + \tilde{\delta} } \right. \\ \label{eq:th-f-0}
	  & \qquad \left. + \, \frac{4 i y_\tau}{y_\tau^2 +1 } \left( 1 - \frac{y_\tau^2}{y_\tau^2 +1} \right) \frac{1 - \eta^b_\tau /2}{\left[ y_\tau^2 + \varw^{-2}
	  \left( y_x^2 + \varv^2 \, y_y^2 \right) + 1 + \tilde{\delta}  \right]^2} \right\} \, .
 \end{align}
  In the limit $\varv \to 0$, the coordinate $y_y$ only enters in the first factor of Eq.~(\ref{eq:th-f-0}).
  However, the loop integral does \emph{not} vanish due to the residue theorem here, since this factor decays too slowly for large $y_y$ in the contributions to $\eta^f_\tau$ in which the frequency derivative act on the multiplicative regulator $ \chi$.
 We switch to the momentum coordinates $y_\pm = y_x \pm y_y$ to obtain 
   \begin{align} \notag
	  \hat{T}^f_\tau (0, \varw,\eta^b_\tau,\tilde{\delta}) &= \lim_{Y \to \infty} \frac{1}{16 \pi^3} \int_{-\infty}^{+\infty} \! d y_\tau \int_{-\infty }^{+ \infty} \! \, d y_+ \, \int_{- Y}^{+ Y} \! d y_- \,\, 
	 \frac{1}{ i y_\tau + y_- } \\ \notag
	 & \quad \times 
	 \left\{  \frac{4 i y_\tau}{\left(y_\tau^2 +1 \right)^2} \left( 1 - 2 \frac{y_\tau^2}{y_\tau^2 + 1 } \right)
		\, \frac{1}{y_\tau^2 + \varw^{-2}
		\left( y_+^2 + y_-^2 \right)/4 + 1 + \tilde{\delta} } \right. \\ \notag
	  & \qquad \left. + \, \frac{4 i y_\tau}{y_\tau^2 +1 } \left( 1 - \frac{y_\tau^2}{y_\tau^2 +1} \right) \frac{1 - \eta^b_\tau /2}{\left[ y_\tau^2 + \varw^{-2}
	  \left( y_+^2 + y_-^2 \right)/4 + 1 + \tilde{\delta}  \right]^2} \right\} \\ \notag
	  &= \frac{\varw}{4 \pi} \left\{ \tilde{\delta}^{-5/2} \left[ \left( 3 + 2 \tilde{\delta} \right) \sinh^{-1} \left( \sqrt{\tilde{\delta}} \right) - 3 \sqrt{ \tilde{\delta} \left( 1 + \tilde{\delta} \right)} \right] \right.\\ \notag
	  & \qquad + \left. \tilde{\delta}^{-3} \left( 1 + \tilde{\delta} \right)^{-1/2}
	  \left[ \left( 3 + \tilde{\delta} \right) \tilde{\delta} - 3 \sqrt{ \tilde{\delta} \left( 1 + \tilde{\delta} \right)} \sinh^{-1} \left( \sqrt{\tilde{\delta}} \right) \right] \left( 1 - \eta^b_\tau /2 \right) \right\} \, ,\\ \label{eq:th-f-0-v0}
	  \hat{T}^f_\tau (0, \varw,\eta^b_\tau,0)  & = \frac{\varw}{5 \pi} \left( \frac{1}{3} + \frac{2-\eta^b_\tau}{4} \right) \, .
  \end{align}
  
  Note that the limit $Y \to \infty$ is to be taken \emph{after} the $y_+$-integral has been performed.
  In the deep infrared limit, Eq.~(\ref{eq:th-f-0-v0}) leads to $ \eta^{f\ast}_\tau = 0.61 $, which is in complete agreement with our numerical results (for pinned flows $\tilde{\delta}=0$).
%

\subsection{Closed fermion loops: Landau damping, bosonic momentum factor}
  \label{sec:landau-damping}
The renormalization of boson propagator factors $\mathcal{A}^b_\tau$ and $\mathcal{A}^b_{xy}$ is determined 
by frequency- and momentum-derivatives of a closed fermionic loop, respectively. Before 
discussing these specific cases in the following, it is useful to write down the general 
expression of the form
 \begin{equation} \label{eq:gen-ferm-bubb}
 \begin{split}
	 L_{m,n}^{(s)} &= \int \! d k \, s \left[ \chi(k_\tau,\Lambda) \right]^{s-1} \, \dot{\chi} (k_\tau,\Lambda) 
	 \left[ i A^f_\tau k_\tau + A^f_x k_x + A^f_y k_y \right]^{-m}
	 \left[ i A^f_\tau k_\tau + A^f_x k_x - A^f_y k_y \right]^{-n} \\
	 &= \frac{ \left( A^f_\tau \Lambda \right)^{2-m-n}}{A^f_x A^f_y} \int \! d y \,
	 s \left[  \frac{y_\tau^2}{y_\tau^2 + 1} \right]^{s-1} \, \frac{2 y_\tau^2}{\left( y_\tau^2 +1 \right)^2}
	 \left[ i  y_\tau + y_x + y_y \right]^{-m}
	 \left[ i  y_\tau + y_x - y_y \right]^{-n} \, .
 \end{split}
 \end{equation}
 In particular, loop integrals of this kind appear on the right-hand side of $ \dot{\mathcal{A}}^b_{xy} $ (with $m=1$, $n=3$, and $s=2$),
 in the four-line fermion ring of Eq.~(\ref{eq:u-flow}) (with $m=n=2$, $s=4$), and in $ \dot{\mathcal{A}}^b_\tau $.
 In that last case, we have $s=2$, $m=1$, and $ n=1,2,3$ depending on whether frequency derivatives act
 on the unregularized propagator or on the regulator $\chi$.
 
 In the first two of these examples, we can straightforwardly apply the residue theorem for the $y_x$-integration.
 The poles then lie in one half of the complex plane and therefore the corresponding integrals vanish for $ \Lambda_\mathrm{UV} \to \infty $
 as in Ref.~\onlinecite{sur14}.
Therefore $\mathcal{A}^b_{xy}$ stops to flow below scales $ \Lambda \ll \Lambda_\mathrm{UV} $ 
consistent with the numerics Fig.~\ref{fig:abxys}.
  For same reason, contributions to $ \dot{\mathcal{A}}^b_\tau $ with $m=2,3$ in Eq.~(\ref{eq:gen-ferm-bubb})
 have to vanish in that regime.
 For the contributions to $\mathcal{A}^b_\tau$ with $m=1$, in which frequency derivatives only act on the regulator, the situation is more subtle:
 In that case, Fubini's theorem does not apply in the limit $ \Lambda_\mathrm{UV} \to \infty $.
 This allows for a finite $ \eta^b_\tau $ in the regime of anomalous logarithmic scaling, where $ \Lambda_\mathrm{UV} $ is virtually infinite.
 As for $\eta^f_\tau$, the power-law contributions to $\eta^b_\tau$ are therefore stabilized by the regulator.
  
    We now prove the scaling relation between the bosonic frequency Landau damping factor 
    and the fermion-boson vertex Eq.~(\ref{eq:screl-g-abt}). With Eq.~(\ref{eq:y_coordinates}), 
  the flow equation for $\mathcal{A}^b_\tau$ in Eq.~(\ref{eq:bose_flow}) becomes  
\begin{equation}
 \eta^b_\tau = \frac{g^2}{\mathcal{A}^b_\tau \Lambda} \, \frac{1}{ A^f_x A^f_y} \, T^b_\tau
 = \tilde{g}^2 \, T^b_\tau \,,
\end{equation}
wherein the dimensionless scaling function
\begin{align} \notag
	T^b_\tau &= - \frac{1}{\pi^3} \lim_{c_x \to \infty}  \int_{-\infty}^{+\infty}d y_\tau  
	\left[ \chi^{(0,0)} (y_\tau,1) \, \chi^{(2,1)} (y_\tau,1) +\chi^{(0,1)} (y_\tau,1) \, \chi^{(2,0)} (y_\tau,1) \right] 
	\\ \notag
 & \quad \times 
  \int_{-c_x}^{+c_x} \! d y_x \,
  \int_{-\infty}^{+\infty} \! d y_y \, \prod_{\pm} \left( i y_\tau + y_x \pm y_y \right)^{-1} \\ \notag
  &= - \frac{1}{\pi^2} \lim_{c_x \to \infty}  \int_{-\infty}^{+\infty}d y_\tau  
	\left[ \chi^{(0,0)} (y_\tau,1) \, \chi^{(2,1)} (y_\tau,1) +\chi^{(0,1)} (y_\tau,1) \, \chi^{(2,0)} (y_\tau,1) \right] 
	\operatorname{sign} \left(y_\tau \right) \, \tan^{-1} \left( c_x/y_\tau \right) \\
  &= \frac{1}{4}
\end{align}
takes on a positive universal value, which is in full agreement with our numerical results.
Here the limit $ \Lambda_\mathrm{UV} \to \infty $ enters as $c_x \to \infty$. We emphasize that it is to be taken \emph{after} the integrals over the
rescaled momentum variables $y_x$ and $y_y$ have been performed.

Due to the scaling relation Eq.~(\ref{eq:screl-fxy}) for the fermions, the product $ A^f_x A^f_y $ does not flow. Therefore, if we require $\eta^b_\tau$
 to be constant, the quotient $ g^2 / (\mathcal{A}^b_\tau \Lambda) $ has to be constant as well, which implies the scaling relation $\eta^b_\tau = 1 + 2 \eta_g$, Eq.~(\ref{eq:screl-g-abt}).
 Moreover, if the value of $\eta^b_\tau$ is universal, also the value of $ \tilde{g}^2 \equiv g^2 / \left( A^f_x A^f_y \mathcal{A}^b_\tau \Lambda \right) $ must be universal, as is indeed the case in our numerical results of Fig.~\ref{fig:g-rat}.
Eq.~(\ref{eq:screl-g-abt}) implies that in the simplified ``Hertz-Millis case'' of ignoring vertex renormalizations ($\partial_\Lambda g=0$),
is characterized by $ \eta^b_\tau = 1 $ at finite scales. 

We now show
how the soft frequency regulator for the fermions achieves this. 
First, let us assume that the dependence of the bosonic self-energy $\Sigma_b$ on a general cutoff $\Lambda$ and on the frequency $\omega$ is governed by the scaling condition 
\begin{equation} \label{eq:sigmab-scale}
 \Sigma_b^{(\Lambda)} \left( \omega , 0 \right) = \Lambda \, \Sigma_b^{(1)} \left( \left| \omega \right| / \Lambda ,0 \right) \, .
\end{equation}
 This relation is fulfilled by the following ansatz
\begin{equation}
\Sigma_b^{(\Lambda)} \left( \omega  , 0 \right) = \left| \omega \right| f  \left( \left| \omega \right| / \Lambda \right) + \Lambda \, g  \left( \left| \omega \right| / \Lambda \right) \, .
\end{equation}
 Requiring regularity of $ \Sigma_b $ in the limit $ \Lambda \to 0 $ implies $ f \left( \left| \omega \right| / \Lambda \right) \to \mathrm{const} $ 
 and  $ g \left( \left| \omega \right| / \Lambda \right) \to \mathrm{const} $, which leads to a Landau damping term in the IR.
 This linear frequency dependence also prevails at large frequencies $ | \omega | \gg \Lambda $. 
 Furthermore, we know the bosonic self-energy to be an analytic function of $\omega$ for $ \Lambda >0 $, such that the non-analyticity of the Landau damping term is only recovered after the
 cutoff has been fully removed. This way, $f(x) $ and $g(x)$ are restricted to be odd and even functions, respectively.
Performing now a gradient expansion of the bosonic self-energy around $ \omega =0 $, the leading-order coefficient is\begin{equation}
 \mathcal{A}^b_\tau = \left. \frac{1}{2} \, \frac{\partial^2}{\partial \omega^2} \, \Sigma_b (\omega, 0) \right|_{\omega =0} \, .
\end{equation}
 Due to the analyticity of the bosonic self-energy at finite scales, one obtains a non-singular coefficient $ \mathcal{A}^b_\tau \propto \frac{1}{\Lambda} $.
 Finally, this corresponds to an anomalous dimension
\begin{equation}
 \eta^b_\tau = - \frac{d \ln \mathcal{A}^b_\tau }{ d \ln \Lambda}  =1 \, .
\end{equation}

 In practice, this result imposes a constraint on the cutoff scheme. In the present context, Landau damping is known to be occur if the fermionic renormalization constants and the Yukawa coupling are kept fixed. A suitable cutoff scheme should leave this behavior intact and therefore fulfill the following conditions:
 (i) For all $\Lambda > 0$, the bosonic self-energy $ \Sigma_b^{(\Lambda)} $ should be an analytic function of the frequency,
 (ii) In perturbation theory, the scaling condition~(\ref{eq:sigmab-scale}) should hold.
 At low scales $ \Lambda \ll \Lambda_\mathrm{UV} $, these conditions are both fulfilled by our soft 
 frequency regulator for the fermions.

 \subsection{Flow of the bosonic mass and critical exponents} 
 \label{sec:flow-mass}

  So far, in the numerics, we have considered the RG flow pinned to criticality, i.e.\ we have set 
  $ \delta_b =0\; \forall \Lambda $ 
  as is also common practice in the field-theoretical renormalization group \cite{MMSS10b,sur14}.
In order to compute the critical exponents also in the vicinity of the critical point, Eq.~(\ref{eq:crit-exp}),
and to derive the fixed-point structure of Eqs.~(\ref{eq:beta-deltat}-\ref{eq:beta-eta}),
we will now include the bosonic mass and self-interaction.
  After rescaling, the flow of the mass Eq.~(\ref{eq:mass-flow}) becomes 
  \begin{equation}
	  \partial_\Lambda \delta_b = \mathcal{A}^b_\tau \Lambda \left[ 
	  3 \tilde{g}^2 - \frac{5 \tilde{u}}{\pi} \left( 1 - \eta^b_\tau /2 \right) \left( 1 + \tilde{\delta} \right)^{-1/2} \right] \, ,
  \end{equation}
  with $ \tilde{u} \equiv u / \left( \mathcal{A}^b_\tau \mathcal{A}^b_{xy} \Lambda \right) $ denoting a rescaled bosonic self-interaction driven by  
%
  \begin{equation}
	  \partial_\Lambda u = \frac{11 u^2}{ 2 \pi \, \mathcal{A}^b_\tau \mathcal{A}^b_{xy} \Lambda^2}  \left( 1 - \eta^b_\tau /2 \right) \left( 1 +\tilde{\delta} \right)^{-3/2}\,,
  \end{equation}
 as follows from Eq.~(\ref{eq:u-flow}). The four-fermion box $\sim g^4$ vanishes in the infrared 
  (see Eq.~(\ref{eq:gen-ferm-bubb})).
  
  Since $\varv$ still decreases logarithmically in the presence of these additional couplings,
  the limit $ \varv \to 0 $ is taken in all diagrams in order to obtain leading power-law contributions to the scaling exponents.
  For the coupling ratios $ \tilde{g} $, $ \varw $ and the rescaled quantities
  \begin{equation}
	  \tilde{\delta} \equiv \frac{\delta_b}{\mathcal{A}^b_\tau \Lambda^2} \, , \quad
	  \tilde{u} \equiv \frac{u}{\mathcal{A}^b_\tau \mathcal{A}^b_{xy} \Lambda} \, ,
  \end{equation}
  we obtain the $\beta$-functions in the infrared shown in Eqs.~(\ref{eq:beta-ut},\ref{eq:beta-eta}).
  The threshold functions $T^{f,b}_{\tau,y}$ appearing here have been written out in the subsections \ref{sec:g-abt-scr},\ref{sec:landau-damping}.

The zeros of the $\beta$-functions of Eqs.~(\ref{eq:beta-deltat}-\ref{eq:beta-eta}) 
correspond to the universal coupling ratios at the RG fixed point in the 
range $ 1 < \eta^b_\tau < 2 $ and $ 4 < \tilde{g} < 8 $.
  Then Eq.~(\ref{eq:beta-ut}) has a physical, namely the trivial solution $\tilde{u}^\ast =0$, and an unphysical one with $ \tilde{u}^\ast < 0 $.
  In other words, the renormalization of the bosonic fields then renders the self-interaction of the bosons irrelevant.
  This effect is induced by the fermionic sector, and hence is not in conflict with the upper critical dimension $ d_\mathrm{c} =4 $ of a purely bosonic theory.
  Together with Eq.~(\ref{eq:beta-eta}), the vanishing of $\tilde{u}$ gives rise to the following simple explicit equation for the rescaled bosonic mass
  \begin{equation}
	  \tilde{\delta}^\ast = \frac{12 \left(\tilde{g}^\ast \right)^2}{ 8 - \left( \tilde{g}^\ast \right)^2} \, ,
  \end{equation}
  which can now be substituted in Eqs.~(\ref{eq:beta-gt}) and (\ref{eq:beta-w}).
  Further, $ \beta_\varw \left(\tilde{g}^\ast,\varw^\ast,\left(\tilde{g}^\ast \right)^2/4,\tilde{\delta}^\ast \right) =0 $ constitutes a quadratic equation
  in $\varw^\ast$ with a highly nontrivial dependence on $\tilde{g}^\ast$.
  One of the solutions is the trivial one, which however is unphysical.
  The other one can as well be substituted in Eq.~(\ref{eq:beta-gt}), and  we then have to solve for $\tilde{g}^\ast$ numerically.
  This way, we find a strange metal fixed point with
  \begin{equation}
	  \tilde{g}^\ast = 4.246 \, , \quad \varw^\ast =15.30 \, , \quad \tilde{\delta}^\ast = 13.57 \, , \quad \tilde{u}^\ast = 0.00 \, .
  \end{equation}
  Note that the scaling relations~(\ref{eq:screl-g-abt}), (\ref{eq:fermbos-screl}) and $ \eta_\delta = \eta^b_\tau -2 $
  only hold up to logarithmic corrections, which are inaccessible in fully rescaled variables and arise 
  as $\varv \to 0$.
  These scaling relations are also reflected in the values of the anomalous dimensions 
  Eq.~(\ref{eq:values-etas}).
  
  Analyzing the stability of this fixed point reveals that three directions are irrelevant.
  The remaining fourth, relevant direction almost coincides with the $\tilde{\delta} $-direction and measures the distance $ \Delta r \equiv r - r_\mathrm{c} $ from criticality.
 This relevant perturbation grows as $ \Delta r \sim \Lambda^\kappa $, where $\kappa$ denotes the corresponding (negative) eigenvalue of the stability matrix.
  Identifying the spin susceptibility as $ \chi = 1/ \delta_b $ and the correlation length as $ \xi = \sqrt{ \mathcal{A}^b_\tau / \delta_b} $,
  we obtain the critical exponents quoted in Eq.~(\ref{eq:crit-exp}) and the associated Grueneisen ratio 
  Eq.~(\ref{eq:grueneisen}).

%
  If we set $ \tilde{\delta} = \tilde{u} =0 $ in the above $\beta$-functions,
  the universal values of the coupling ratios and associated anomalous exponents obtained as zeros of these $\beta$-functions are in full agreement with those found in the numerics Sec.~\ref{sec:numerics}. 

\section{Comparison with previous work}
\label{sec:comparison}

\subsection{Abanov-Chubukov (2000), $1/N_f$-expansion to 1-loop}
In a classic paper \cite{abanov00}, Abanov and Chubukov resummed logarithmic 
singularities appearing in vertex and self-energy one-loop corrections
for the spin-fermion model. As the Bose propagator, the Landau-damped 
form from the bubble with bare fermions is used. The graphs 
are selected based on a $1/N_f$ expansion with $N_f$ the number of 
fermion flavours, which was later shown to be inapplicable to the problem
\cite{MMSS10b}. The asymptotically, locally nesting of Fermi surfaces 
directly at the hot spot was predicted and the critical exponents 
for the electron and spin-wave propagators were computed. 
The anomalous logarithmic corrections from asymptotic nesting on top of the 
power-laws were ``left standing'' and not discussed much further.

Next to our different RG scheme (explained in Sec.~\ref{sec:RG}) 
our work goes beyond that of Abanov and Chubukov in that we include 
the mass of the bosonic order parameter field ($\tilde{\delta}$) and 
bosonic quartic coupling ($\tilde{u}$) into the RG flow (Eqs.~(\ref{eq:beta-deltat}-\ref{eq:beta-eta})).
While the $\tilde{u}^\ast=0$ at the fixed-point, $\delta^\ast\neq 0$ and this 
reduces for example the power-law divergence in the vertex significantly:
$\eta^\ast_g = 0.03$ in the present paper versus $\eta^{A\&C}_g = 0.125$.
Furthermore, using scaling of the $\delta$ in the vicinity of the critical point, 
we provide the correlation length and susceptibility exponents ($\nu$ and $\gamma$), 
which were not computed in Ref.~\onlinecite{abanov00}.
But perhaps most significantly, we cast the RG flow in a form using dimensionless 
variables, in which the anomalous logarithms due to asymptotic nesting disappear 
and it is possible to take the $\varv \rightarrow 0$ limit explicitly. Our form 
of the RG equations Eqs.~(\ref{eq:beta-deltat}-\ref{eq:beta-eta}) and the truncation can now 
systematically be expanded, keeping for example momentum and/or-frequency dependencies of the couplings.

\subsection{Metlitski-Sachdev (2010), $1/N_f$-expansion to 2-loop}
Everything that applies to the comparison of the present work to Abanov and Chubukov 
essentially also applies to the one-loop graphs computed in the $1/N_f$ 
expansion by Metlitski and Sachdev \cite{MMSS10b}. However, the 
logarithmic singularities from asymptotic nesting were taken more seriously 
in Ref.~\cite{MMSS10b} and interpreted as divergences in the critical exponents and 
a breakdown of the RG. Again our set of dimensionless variables 
used to solve the RG equations and the soft frequency technique made the $\varv \rightarrow 0$ limit 
transparent and possible in both, the analytical form Eqs.~(\ref{eq:beta-deltat}-\ref{eq:beta-eta})
as well as the numerics (Fig.~\ref{fig:flow-v}).

Ref.~\cite{MMSS10b} found additional severe $\varv \rightarrow 0$ divergences in the 2-loop corrections 
to the Bose propagator. While it is possible to include certain 2-loop effects into the functional RG framework 
(see Refs.~\onlinecite{katanin09, eberlein14} for fermions), at this point in time we cannot assess whether similar issues would appear within 
our soft frequency regulator technique.  What can be said philosophically that, 
even if we had a direct 2-loop graph in our RG, the frequency regulator would impose phase 
space constraints different from the direct, ``straight-up'' loop integrations, which mix all energy shells more freely.
Moreover, we would not be using the overdamped and non-local RPA propagator of the $1/N_f$ expansion; 
within our RG we use propagators of a local field theory in all the graphs (i.e. in the Bose propagator 
we use $\mathcal{A}^b_\tau \omega^2$ instead of the $|\omega|$ used in the $1/N_f$ expansion).
This last feature is also true for the $\epsilon$-expansion by Sur and Lee \cite{sur14}.

\subsection{Sur-Lee (2015), $\epsilon = 3-d$ expansion to $O(\epsilon)$}
\label{subsec:surlee}
  Our strange metal fixed point shares the qualitative features with the one obtained by Sur and Lee in $ 3 - \epsilon $ dimensions \cite{sur14}.
  In both cases, $\varv$ flows to zero logarithmically and $ \varw $ takes on a nonzero universal value at the fixed point.
  The dynamical exponent of Sur and Lee extrapolated to $\epsilon = 1$ takes on the 
value $z_{\epsilon =1} = 1.83$ which is larger than our value following from the comparable 
additive definition below Eq.~(\ref{eq:dynamical}): $z=1.53$.
  While also a rescaled Yukawa coupling approaches a finite universal value in either case, our rescaling differs 
  from Ref.~\onlinecite{sur14}.
Sur and Lee used the coupling ratio
 %
	$\lambda \equiv \frac{ \tilde{g}^2 A^f_\tau }{ A^f_y }$
%
as the effective coupling strength between the electrons and spin waves.
This coupling ratio naturally arises from their isotropic rescaling.
In the present analysis of the two-dimensional case, this quantity diverges with
$ \eta^\ast_\lambda =\eta^{b\ast}_\tau /2 - 2 \eta^{f\ast}_y $, while it flows to a universal 
constant in the $\epsilon$ expansion. 
This may indicate a possible breakdown of the $\epsilon$-expansion to $ \epsilon =1 $ in 
that the $\epsilon$-expansion can only pick-up logarithmic singularities (i.e. $1/\epsilon$ poles) 
and will miss for example linear power-law scaling.  Furthermore, in the rescaled $\beta$-functions, we use a  (purely cosmetic) anisotropic rescaling of the boson propagator, while keeping the fermions fully two-dimensional.
\begin{table}[t]
	\begin{tabular}{|c| >{$}c<{$} | >{$}c<{$} | >{$}c<{$} | >{$}c<{$} | >{$}c<{$} | >{$}c<{$} | >{$}c<{$} |  >{$}c<{$} |}
		\hline
		Sur and Lee & \mathcal{A}_1& \mathcal{A}_2& \mathcal{A}_3& \mathcal{A}_4& \mathcal{A}_5& \mathcal{A}_6 & \mathcal{A}_7, \mathcal{A}_8 &  \\
		\hline
		this work & A^f_\tau & A^f_x & A^f_y & \mathcal{A}^b_\tau & \mathcal{A}^b_{xy} & g & u & \delta_b \\
		\hline
	\end{tabular}
	\caption{Correspondence between the counter terms of Ref.~\onlinecite{sur14} with strengths $ \mathcal{A}_i$ and the running couplings of the present work.
	In the former, $ \mathcal{A}_7 = \mathcal{A}_8 $ for the $\phi^4$ term in the (physical) case of an SU(2) spin symmetry.
	Flows of our bosonic mass $\delta_b$ are not included within the epsilon-expansion approach of Ref.~\onlinecite{sur14}.}
	\label{tab:correspondence}
\end{table}

 A convenient feature of the $\epsilon$-expansion and the embedding of the fermionic dispersion is that 
  the two-dimensional metal becomes semimetallic in $d>2$ with a vanishing density 
  of states at the Fermi level. With our analysis in $d=2$ directly, the electronic density of states
 $ \rho (\omega) \sim \Lambda_\mathrm{UV} / ( A^f_x A^f_y ) $ remains constant 
 during the flow. 
 
We believe the leading contributions of the $\epsilon$-expansion 
 are captured within our analysis and it will be interesting to compare 
 higher-order in $\epsilon$ computations to more extended truncations of
 Polchinski-type flow equations with truly functional frequency 
 and/or field dependencies.

\section{Conclusions}

In this paper, we derived the universal low-energy asymptotics of two-dimensional 
electronic metals at the onset of antiferromagnetism at zero temperature. 
By systematically decimating 
energy shells from high to lowest energies along a continuous 
flow parameter $\Lambda$ for both, the electrons and 
the collective spin-waves, we obtained (i) a ``strange metal'' fixed-point characterized by 
non-Fermi liquid behavior of the electrons at the hot spots, (ii) the associated critical exponents 
in the vicinity of the critical point, (iii) analytic expressions for universal amplitudes and scaling 
functions, and (iv) non-universal features such as crossover scales from 
the numerical solution of coupled renormalization group equations for the effective action. 
We showed by explicit computation
that this fixed-point is accessible from an initially weakly coupled model and strong correlations 
build up only for shrinking energy shells as $\Lambda$ is lowered.

Moreover, we were able to demonstrate the truly universal nature of this
antiferromagnetic strange metal, by obtaining the same theory in the limit 
$\Lambda\rightarrow 0$ for a range of initial parameters such as different Fermi surface angles 
at the hot spots. We also computed the energy range of an intermediate crossover regime, 
during which the flow ``looses its memory'' about the initial conditions. This implies that 
whenever a commensurate spin-density wave quantum critical point with a non-nested 
hot spot geometry is invoked for a two-dimensional 
material across the cuprate, pnictide, and heavy-fermion families, the behavior at lowest 
temperatures falls into the same universality class (irrespective of for example details of the 
``bare'' Fermi surface angles) and should be governed by the same critical exponents. 

The physical results and techniques developed in this paper can now form a basis for:  
(i) systematic, energy-resolved investigations of competing instabilities such 
as superconductivity and charge order formation close to antiferromagnetic quantum critical 
points,
 (ii) quantitatively accurate computations of the critical exponents of this universality class directly 
in spatial $d=2$, and 
(iii) ``UV-completions'' of the antiferromagnetic strange metal via 
a feeding the flow 
material parameters from ab-initio techniques to quantitatively resolve the (non-universal)
validity range of low-energy models and crossover scales for specific compounds.

\acknowledgments
We are grateful to Aavishkar Patel and Subir Sachdev for many discussions and collaboration 
on related topics. We further thank Holger Gies, Achim Rosch, Manfred Salmhofer, and Peter W\"olfle for helpful
discussions. This research was supported by the Leibniz prize of A.\ Rosch.
The numerical calculations have been carried out on the CHEOPS cluster at the University of Cologne and 
on a corresponding program on the RWTH cluster in Aachen (thanks to Carsten Honerkamp).

\appendix

 \section{Vanishing scale derivatives of the fermionic propagator}
 \label{sec:dyn_ferm}

 In our calculations, we have neglected the ``dynamical'' scale derivatives in the fermionic single-scale propagator, i.e.\ we have omitted the second term in
  \begin{equation}
	  S^{(\pm)}_f =  \frac{G^{(\pm),R}_f \dot{\chi}}{ \chi}
	  + \frac{\chi^{-1} -1}{\Lambda} \left( G^{(\pm),R}_f \right)^2 
	  \left[ \eta^f_\tau i A^f_\tau k_\tau - \eta^f_y \left( A^f_x k_x \mp A^f_y k_y \right) \right] \, .
  \end{equation}
  In the following, we show that this term leaves the universal deep IR properties unaffected.

  To this end, let us first consider $\eta^f_y = 3 \tilde{g}^2 \mathcal{T}^f_y $
  with the threshold function $\mathcal{T}^f_y = T^f_y + \mathcal{F}^f_y$, which consists of the contributions $T^f_y$ given in Eq.~(\ref{eq:tfy})
  and the fermionic dynamical part
  \begin{align} \notag
  \mathcal{F}^f_y &= \int_y
 	 \frac{1}{\left( i y_\tau + y_x - y_y \right)^3} 
	 \frac{2 y_\tau^2}{\left(y_\tau^2 +1 \right)^2} \, \frac{1}{y_\tau^2 + \varw^{-2}
	 \left( y_x^2 + \varv^2 \, y_y^2 \right) + 1 + \tilde{\delta} }
	 \left[ i \eta^f_\tau y_\tau - \eta^f_y \left( y_x + y_y \right) \right] \\ \label{eq:Ffy}
   &\quad - \int_y
 	 \frac{1}{\left( i y_\tau + y_x - y_y \right)^2} 
	 \frac{ y_\tau^2}{\left(y_\tau^2 +1 \right)^2} \, \frac{1}{y_\tau^2 + \varw^{-2}
	 \left( y_x^2 + \varv^2 \, y_y^2 \right) + 1 + \tilde{\delta} } \, \eta^f_y \, .
 \end{align}
  Let us now perform the $y_y$ integration first.
  In the limit $\varv \to 0$, we then find that $\mathcal{F}^f_y$ vanishes according to the residue theorem.
  Hence, also a fully dynamical RG flow will yield $\eta^{f\ast}_y = - \eta^{f\ast}_x = 0$, and we only have to take the dynamical scale derivative $\propto \eta^f_\tau$
  into account in the following.

  Again by virtue of the residue theorem, the dynamical fermionic contributions vanish in $\eta^b_\delta $, $\eta^b_\tau$, and $\eta^b_{xy}$.
  In the strict deep IR limit,
  this also holds for $\eta^f_\tau = 3 \tilde{g}^2 \left( T^f_\tau + \mathcal{F}^f_\tau \right)$, where
 \begin{align}  \notag
	 \mathcal{F}^f_\tau &= - \int_y
 	 \frac{1}{\left( i y_\tau + y_x - y_y \right)^3} \,
	 \frac{2 y_\tau^2}{\left(y_\tau^2 +1 \right)^2} \, \frac{i \eta^f_\tau y_\tau - \eta^f_y \left( y_x + y_y \right) }{y_\tau^2 + \varw^{-2}
	 \left( y_x^2 + \varv^2 \, y_y^2 \right) + 1 + \tilde{\delta} } \\ \notag
	 & \quad - \int_y \frac{1}{ \left( i y_\tau + y_x - y_y \right)^2 } \,
	 \frac{2 i y_\tau}{\left(y_\tau^2 +1 \right)^2} \left( 1 - 2 \frac{y_\tau^2}{y_\tau^2 + 1 } \right)
		\, \frac{i \eta^f_\tau y_\tau - \eta^f_y \left( y_x + y_y \right)}{y_\tau^2 + \varw^{-2}
		\left( y_x^2 + \varv^2 \, y_y^2 \right) + 1 + \tilde{\delta} } \\
	 & \quad + \int_y \frac{1}{ \left( i y_\tau + y_x - y_y \right)^2 } \, \frac{y_\tau^2}{\left(y_\tau^2 +1 \right)^2}
		\, \frac{ \eta^f_\tau }{y_\tau^2 + \varw^{-2}
		\left( y_x^2 + \varv^2 \, y_y^2 \right) + 1 + \tilde{\delta} } \, ,
 \end{align}
 which then vanishes for $\varv \to 0$.

  However, this argument does not apply for $\eta_g$, and one may wonder, whether the scaling relation~(\ref{eq:screl-g-abt}) still holds in a fully dynamical RG calculation.
  Indeed it does hold, which becomes evident after the momentum integrals in the corresponding threshold function $T_g$ have been carried out.
  In order to show this, let us again look at the fermionic dynamic contribution
 \begin{align} 
 \label{eq:Fg} \notag
 \mathcal{F}_g &= \frac{1}{8 \pi^3} \int_{-\infty}^{+\infty}d y_\tau  \int_{-\infty}^{+\infty} d y_x  
	 \int_{-\infty}^{+\infty} d y_y \,\, 
	 \frac{2 y_\tau^4}{\left(y_\tau^2 +1 \right)^3} \, \frac{i \eta^f_\tau y_\tau - \eta^f_y \left( y_x + y_y \right)}{y_\tau^2 + \varw^{-2}
	 \left( y_x^2 + \varv^2 \, y_y^2 \right) + 1 + \tilde{\delta} }
	 \\ & \quad \times 
	 \frac{1}{ i y_\tau + y_x + y_y} \, \frac{1}{ \left( i y_\tau + y_x - y_y \right)^2}
 \end{align}
 to the corresponding threshold function, which enters in $\eta_g = \tilde{g}^2 \left( T_g + \mathcal{F}_g \right) $
 with $ T_g $ as in Eq.~(\ref{eq:tg}). Since we have already shown that $ \eta^{f\ast}_y =0 $,
 let us consider the limit $\varv, \eta^f_y \to 0$.  If we then perform the momentum integrals, we obtain \begin{equation}
	 \mathcal{F}_g \to - \frac{\varw}{8 \pi} \int_{-\infty}^{+\infty} \! d y_\tau \,
	 \frac{\eta^f_\tau y_\tau}{\sqrt{1 + y_\tau^2 +\tilde{\delta}} \,
	 \left( \left| y_\tau \right| + \varw \sqrt{1 + y_\tau^2 +\tilde{\delta}} \right)^2} \,
         \frac{y_\tau^4}{\left(y_\tau^2 +1 \right)^3} =0  \, .
 \end{equation}
  We observe that the integrand of the remaining frequency integral is then antisymmetric in the limit $\varv \to 0$,
  and consequently also these dynamical contributions vanish.

  Summarizing, we therefore can state that a fully dynamical flow approaches the very same fixed point as above, while the precise RG trajectory in parameter space differs from the RG flows
  analyzed in this work. The fermionic dynamical contributions can be expected to have their strongest 
  impact at $\Lambda \lesssim\Lambda_{UV} $, where the finiteness of the UV cutoff plays a role.
  
 \section{Explicit expressions for the coupled flow equations} \label{sec:explicit-flow}

In this Appendix, we give explicit expressions for the diagrams on the right-hand sides of the six flow equations (\ref{eq:yuk_flow}-\ref{eq:fermi_flow}).
 For simplicity, we only have only constrained the $k_x$ momentum coordinate, but not the $k_y$ coordinate, i.e.\ 
 \begin{equation}
	 \int \! d k = \frac{1}{8 \pi^3} \int_{- \infty}^{+\infty} \! d k_\tau \int_{- \Lambda_\mathrm{UV}}^{+\Lambda_\mathrm{UV}} \! d k_x \int_{- \infty}^{+\infty} \! d k_y \; .
 \end{equation}
 In all diagrams, the $k_y$ integral can then be performed analytically using contour techniques.
 In diagrams where this still leads to numerically tractable expressions, we also calculate one or both of the remaining integrals analytically.

\textit{Yukawa vertex:}
 For the increment of the Yukawa coupling $g$, the $k_y$-integral is performed analytically over the whole real axis
 and the remaining two integrals are calculated using an adaptive multidimensional
 quadrature routine. \cite{dcuhre}
 More precisely, we have
 \begin{equation} \label{eq:num-int-yuk}
 \partial_\Lambda g^2 = \frac{g^4}{4 \pi^3} \int_{-\infty}^{+\infty} \! d k_\tau \int_{- \Lambda_\mathrm{UV}}^{+ \Lambda_\mathrm{UV}} \! d k_x \, F_g (k_\tau,k_x)
\end{equation}
 with the integrand
\begin{align} \notag
F_g (k_\tau,k_x) &=
\frac{\pi  k_\tau^4 \Lambda \Theta (k_\tau ) }{2 {\mathcal{A}^b_{xy}}^2 {A^f_y}^2
	       {p_b}^3 ({p_b}-i {p^{(-)}_f})^2 ({p_b}+i {p^{(+)}_f})^2 ({p^{(-)}_f}-{p^{(+)}_f}) \left(k_\tau^2+\Lambda^2\right)^3} \\ \notag
	       & \quad \times \left\{ -4 i {p_b}^3 \left[\left(k_\tau^2+\Lambda^2\right) (6 {\mathcal{A}^b_\tau}+{\dot{\mathcal{A}}^b_\tau} \Lambda)+4
   {\mathcal{A}^b_{xy}} \left({p^{(-)}_f} {p^{(+)}_f}+{k_x}^2\right)+4 \delta_b \right] \right. \\ \notag
   & \qquad  -{p_b}^2 ({p^{(-)}_f}-{p^{(+)}_f}) \left[5 \left(k_\tau^2+\Lambda^2\right) (6 {\mathcal{A}^b_\tau}+{\dot{\mathcal{A}}^b_\tau} \Lambda)+4
      {\mathcal{A}^b_{xy}} \left({p^{(-)}_f} {p^{(+)}_f}+5 {k_x}^2\right)+20 \delta_b \right] \\ \notag
    & \qquad  +2 i {p_b} ({p^{(-)}_f}-{p^{(+)}_f})^2 \left[6 {\mathcal{A}^b_\tau} \left(k_\tau^2+\Lambda^2\right)+4 {\mathcal{A}^b_{xy}}
    {k_x}^2+{\dot{\mathcal{A}}^b_\tau} k_\tau^2 \Lambda+{\dot{\mathcal{A}}^b_\tau} \Lambda^3+4 \delta_b \right] \\ \notag
    & \qquad -{p^{(-)}_f} {p^{(+)}_f} ({p^{(-)}_f}-{p^{(+)}_f}) \left[6 {\mathcal{A}^b_\tau} \left(k_\tau^2+\Lambda^2\right)+4
    {\mathcal{A}^b_{xy}} {k_x}^2+{\dot{\mathcal{A}}^b_\tau} k_\tau^2 \Lambda+{\dot{\mathcal{A}}^b_\tau} \Lambda^3+4 \delta_b\right]\\ \notag
	 & \qquad \left. +4 {\mathcal{A}^b_{xy}} {p_b}^4 ({p^{(+)}_f}-{p^{(-)}_f}) \right\} \\ \notag
	       & \quad + \frac{\pi  k_\tau^4 \Lambda \Theta (-k_\tau) }{2 {\mathcal{A}^b_{xy}}^2 {A^f_y}^2 {p_b}^3 ({p_b}+i {p^{(-)}_f})^2 ({p_b}-i
			         {p^{(+)}_f})^2 ({p^{(-)}_f}-{p^{(+)}_f}) \left(k_\tau^2+\Lambda^2\right)^3} \\ \notag
				 & \quad \times \left\{4 i {p_b}^3 \left[\left(k_\tau^2+\Lambda^2\right) (6 {\mathcal{A}^b_\tau}+{\dot{\mathcal{A}}^b_\tau} \Lambda)+4 {\mathcal{A}^b_{xy}} \left({p^{(-)}_f} {p^{(+)}_f}+{k_x}^2\right)+4 \delta_b
		     \right] \right. \\ \notag
		     & \qquad  -{p_b}^2 ({p^{(-)}_f}-{p^{(+)}_f}) \left[5 \left(k_\tau^2+\Lambda^2\right) (6 {\mathcal{A}^b_\tau}+{\dot{\mathcal{A}}^b_\tau} \Lambda)+4 {\mathcal{A}^b_{xy}} \left({p^{(-)}_f} {p^{(+)}_f}+5 {k_x}^2\right)+20
		        \delta_b\right]\\ \notag
			& \qquad -2 i {p_b} ({p^{(-)}_f}-{p^{(+)}_f})^2 \left[6 {\mathcal{A}^b_\tau} \left(k_\tau^2+\Lambda^2\right)+4 {\mathcal{A}^b_{xy}} {k_x}^2+{\dot{\mathcal{A}}^b_\tau} k_\tau^2 \Lambda+{\dot{\mathcal{A}}^b_\tau}
			   \Lambda^3+4 \delta_b\right] \\ \notag
			   & \qquad -{p^{(-)}_f} {p^{(+)}_f} ({p^{(-)}_f}-{p^{(+)}_f}) \left[6 {\mathcal{A}^b_\tau} \left(k_\tau^2+\Lambda^2\right)+4 {\mathcal{A}^b_{xy}} {k_x}^2+{\dot{\mathcal{A}}^b_\tau} k_\tau^2
	\Lambda+{\dot{\mathcal{A}}^b_\tau} \Lambda^3+4 \delta_b\right]\\ 
	& \qquad \left. +4 {\mathcal{A}^b_{xy}} {p_b}^4 ({p^{(+)}_f}-{p^{(-)}_f})\right\} \, ,
\end{align}
where the poles of the fermionic lines in $k_y$ are given by
\begin{equation}
	p_f^{(\pm)} = \mp \frac{i A^f_\tau k_\tau + A^f_x k_x}{A^f_y}
\end{equation}
and those of the bosonic lines by $ \pm i p_b $ with
\begin{equation}
	p_b = \frac{\sqrt{\delta_b+ \mathcal{A}^b_\tau \left( k_\tau^2 + \Lambda^2 \right) + \mathcal{A}^b_{xy} k_x^2 }}{\sqrt{\mathcal{A}^b_{xy}}} \, .
\end{equation}
Note that $ \mathcal{A}^b_\tau $ must be calculated from Eq.~(\ref{eq:fbtau}) prior to the vertex correction diagram.

\textit{Fermion self-energy:}
The fermion-self energy flows according to:
 \begin{equation} \label{eq:num-int-afx}
 \partial_\Lambda A^f_x = \frac{3 g^2}{8\pi^3} \int_{-\infty}^{+\infty} \! d k_\tau \int_{- \Lambda_\mathrm{UV}}^{+ \Lambda_\mathrm{UV}} \! d k_x \, F^f_x (k_\tau,k_x)
\end{equation}
with
\begin{align} \notag
F^f_x (k_\tau,k_x) &=
\frac{\pi  {A^f_x} k_\tau^2 \Lambda \Theta (-k_\tau)}{2 {\mathcal{A}^b_{xy}}^2 {A^f_y}^2 {p_b}^3 ({p_b}+i {p^{(-)}_f})^3 \left(k_\tau^2+\Lambda^2\right)^2}\\ \notag
& \quad \times \left\{ ( 3 {p_b}+i {p^{(-)}_f}) \left[\left(k_\tau^2+\Lambda^2\right) (4 {\mathcal{A}^b_\tau}+{\dot{\mathcal{A}}^b_\tau} \Lambda)+2 \delta_b
   \right] \right. \\ \notag
   & \qquad \left. +{\mathcal{A}^b_{xy}} \left(-2 {p_b}^3+2 i {p_b}^2 {p^{(-)}_f}+6 {p_b} {k_x}^2+2 i {p^{(-)}_f} {k_x}^2\right)\right\}\\ \notag
   & \quad +\frac{\pi  {A^f_x} k_\tau^2 \Lambda \Theta (k_\tau) }{2 {\mathcal{A}^b_{xy}}^2 {A^f_y}^2 {p_b}^3 ({p_b}-i {p^{(-)}_f})^3 \left(k_\tau^2+\Lambda^2\right)^2} \\ \notag
   & \quad \times \left\{ (3 {p_b}-i {p^{(-)}_f}) \left[\left(k_\tau^2+\Lambda^2\right)
   (4 {\mathcal{A}^b_\tau}+{\dot{\mathcal{A}}^b_\tau} \Lambda)+2 \delta_b\right] \right. \\
   & \qquad \left. -2 {\mathcal{A}^b_{xy}} \left({p_b}^3+i {p_b}^2 {p^{(-)}_f}-3 {p_b} {k_x}^2+i {p^{(-)}_f} {k_x}^2\right)\right\} \, ,
\end{align}
 and, by construction
\begin{equation}
	\partial_\Lambda A^f_y (k_\tau,k_x) = - \frac{A^f_y}{A^f_x} \partial_\Lambda A^f_x (k_\tau,k_x) \, .
\end{equation}
Finally,
\begin{equation} \label{eq:num-int-af0}
 \partial_\Lambda A^f_\tau = \frac{3 g^2}{8\pi^3} \int_{-\infty}^{+\infty} \! d k_\tau \int_{- \Lambda_\mathrm{UV}}^{+ \Lambda_\mathrm{UV}} \! d k_x \, F^f_\tau (k_\tau,k_x)
\end{equation}
 with
\begin{align} \notag
F^f_\tau (k_\tau,k_x) &=
\frac{\pi  k_\tau \Lambda \Theta (-k_\tau) }{2 {\mathcal{A}^b_{xy}}^2 {A^f_y}^2 {p_b}^3 ({p_b}+i {p^{(-)}_f})^3 \left(k_\tau^2+\Lambda^2\right)^3} \\ \notag
& \quad \times \left( 4 {A^f_y} {p_b}^2 \left[-\left(k_\tau^2+\Lambda^2\right) \left(2 {\mathcal{A}^b_\tau} k_\tau^2-4 {\mathcal{A}^b_\tau} \Lambda^2-{\dot{\mathcal{A}}^b_\tau}
   \Lambda^3\right) \right. \right. \\ \notag
   & \qquad \quad \left. -2 i {\mathcal{A}^b_{xy}} \left(k_\tau^2-\Lambda^2\right) \left({p_b} {p^{(-)}_f}-i {k_x}^2\right)-2 \delta_b \left(k_\tau^2-\Lambda^2\right)\right] \\ \notag
   & \qquad  +(3 {p_b}+i {p^{(-)}_f}) \left\{ A^f_\tau k_\tau \left(8 {\mathcal{A}^b_\tau} k_\tau^4+4 {\mathcal{A}^b_\tau} k_\tau^2 \Lambda^2-4 {\mathcal{A}^b_\tau} \Lambda^4+{\dot{\mathcal{A}}^b_\tau} k_\tau^4 \Lambda-{\dot{\mathcal{A}}^b_\tau} \Lambda^5+6
         \delta_b k_\tau^2-2 \delta_b \Lambda^2\right) \right. \\ \notag
	 & \qquad \quad	\left. -2 i {A^f_x} {k_x} \left[\left(k_\tau^2+\Lambda^2\right) \left(2 {\mathcal{A}^b_\tau} k_\tau^2-4 {\mathcal{A}^b_\tau} \Lambda^2-{\dot{\mathcal{A}}^b_\tau}
\Lambda^3\right)+2 \delta_b \left(k_\tau^2-\Lambda^2\right)\right]\right\} \\ \notag
	    & \qquad \left. +2 {\mathcal{A}^b_{xy}} \left(i {p_b}^3+{p_b}^2 {p^{(-)}_f}-3 i {p_b} {k_x}^2+{p^{(-)}_f} {k_x}^2\right)
	    \left[2 {A^f_x} {k_x} \left(k_\tau^2-\Lambda^2\right)+i {A^f_\tau} \left(3 k_\tau^3-k_\tau \Lambda^2\right)\right] \right) \\ \notag
 & \quad  +\frac{\pi  k_\tau \Lambda \Theta (k_\tau) }{2 {\mathcal{A}^b_{xy}}^2 {A^f_y}^2 {p_b}^3 ({p_b}-i {p^{(-)}_f})^3 \left(k_\tau^2+\Lambda^2\right)^3} \\ \notag
 & \quad \times \left( 4 {A^f_y} {p_b}^2 \left[\left(k_\tau^2+\Lambda^2\right)
	 \left(2 {\mathcal{A}^b_\tau} k_\tau^2-4 {\mathcal{A}^b_\tau} \Lambda^2-{\dot{\mathcal{A}}^b_\tau} \Lambda^3\right) \right. \right. \\ \notag
& \qquad \quad \left.  +2 {\mathcal{A}^b_{xy}} \left(k_\tau^2-\Lambda^2\right) \left({k_x}^2-i {p_b} {p^{(-)}_f}\right)+2 \delta_b
		        \left(k_\tau^2-\Lambda^2\right)\right] \\ \notag
 & \qquad +(3 {p_b}-i {p^{(-)}_f}) \left\{ {A^f_\tau} k_\tau \left(8 {\mathcal{A}^b_\tau} k_\tau^4+4 {\mathcal{A}^b_\tau} k_\tau^2 \Lambda^2-4 {\mathcal{A}^b_\tau}
 \Lambda^4+{\dot{\mathcal{A}}^b_\tau} k_\tau^4 \Lambda-{\dot{\mathcal{A}}^b_\tau} \Lambda^5+6 \delta_b k_\tau^2-2 \delta_b \Lambda^2\right) \right.\\ \notag
 & \qquad \quad \left. -2 i {A^f_x} {k_x} \left[\left(k_\tau^2+\Lambda^2\right)
 \left(2 {\mathcal{A}^b_\tau} k_\tau^2-4 {\mathcal{A}^b_\tau} \Lambda^2-{\dot{\mathcal{A}}^b_\tau} \Lambda^3\right)+2 \delta_b \left(k_\tau^2-\Lambda^2\right)\right] \right\} \\
 & \qquad \left.  +2 {\mathcal{A}^b_{xy}} \left({p_b}^3+i {p_b}^2 {p^{(-)}_f}-3 {p_b} {k_x}^2+i {p^{(-)}_f} {k_x}^2\right) \left[{A^f_\tau} \left(k_\tau \Lambda^2-3 k_\tau^3\right)+2 i {A^f_x} {k_x}
			 \left(k_\tau^2-\Lambda^2\right)\right]\right) \, .
\end{align}

\textit{Boson self-energy:}
 For the bosonic self-energy, also the frequency integral is performed analytically.
 The flow of the frequency term is governed by
 \begin{equation} \label{eq:num-int-ab0}
	 \partial_\Lambda \mathcal{A}^b_\tau = \frac{ g^2}{\pi^3} \int_{- \Lambda_\mathrm{UV}}^{+ \Lambda_\mathrm{UV}} \! d k_x \, F^b_\tau (k_x) \, ,
 \end{equation}
  where
\begin{align} \notag
F^b_\tau (k_x) & =
\frac{\pi  {A^f_\tau}^3 \Lambda}{3 {A^f_y} \left({A^f_x}^2 {k_x}^2-{A^f_\tau}^2 \Lambda^2\right)^5} \\ \notag
& \quad \times \left[5 {A^f_\tau}^6 \Lambda^6-9 {A^f_\tau}^4 {A^f_x}^2 {k_x}^2 \Lambda^4-45 {A^f_\tau}^2 {A^f_x}^4 {k_x}^4 \Lambda^2 \right. \\ \label{eq:fbtau}
	 & \qquad \left. -6 {A^f_x}^2 {k_x}^2
   \left(-2 {A^f_\tau}^4 \Lambda^4+7 {A^f_\tau}^2 {A^f_x}^2 {k_x}^2 \Lambda^2+3 {A^f_x}^4 {k_x}^4\right) \ln \left(\frac{{A^f_x}^2 {k_x}^2}{{A^f_\tau}^2
      \Lambda^2}\right)+49 {A^f_x}^6 {k_x}^6\right] \, .
\end{align}
The remaining integral is calculated numerically using the NAG quadrature routine \texttt{d01sjc}. \cite{nag}
 In addition, we have to expand this integrand around $ k_x =0 $ and $k_x = A^f_\tau \Lambda/A^f_x $ in order to avoid arithmetic overflow in our calculations.

 In the flow equation
 \begin{equation}
	 \partial_\Lambda \mathcal{A}^b_{xy} = \frac{ g^2}{2 \pi^3} \, F^b_{xy}
 \end{equation}
 for the bosonic dispersion all integrals can conveniently be performed analytically, which yields
\begin{align} \notag
F^b_{xy} &=
 -\frac{2 \pi  {A^f_\tau} \Lambda_\mathrm{UV} \Lambda \left({A^f_x}^2+{A^f_y}^2\right) }{{A^f_y} \left({A^f_x}^2 \Lambda_\mathrm{UV}^2-{A^f_\tau}^2 \Lambda^2\right)^4} \\
 & \quad \times \left[{A^f_\tau}^4 \Lambda^4+4 {A^f_\tau}^2 {A^f_x}^2 \Lambda_\mathrm{UV}^2 \Lambda^2+4 \left(2 {A^f_\tau}^2
   {A^f_x}^2 \Lambda_\mathrm{UV}^2 \Lambda^2+{A^f_x}^4 \Lambda_\mathrm{UV}^4\right) \ln \left(\frac{{A^f_x} \Lambda_\mathrm{UV}}{{A^f_\tau} \Lambda}\right)-5 {A^f_x}^4 \Lambda_\mathrm{UV}^4\right] \, .
\end{align}
Also here, we have to resort to an expansion at scales around 
$\Lambda = A^f_x \Lambda_\mathrm{UV} / A^f_\tau$. 

\section{Numerical techniques} \label{sec:numerical-setup}

  In most contributions to the right-hand-sides of the flow equations, 
  an analytical calculation of all three (frequency and momentum) loop integrals is 
  not possible.
  We therefore have to resort to numerical methods to evaluate the remaining one- or two-dimensional integrals.
  Since we expect the corresponding integrands to develop increasingly sharp peaks around zero frequency and momenta with decreasing scale $\Lambda$,
  this represents a nontrivial task.
  In particular, we have to be able to solve the RG flow down to relatively low scales in order to reach the scaling regime.
  This is accomplished by a substitution of the integration variables $ k_i \to K_i = \log k_i $, which is applied after the integrands have been symmetrized in the remaining integration variables.
  We perform this substitution in all integration variables in the numerical quadratures.
  The peaks of the integrand then have a much larger width, which is more easy to sample, and are shifted under a lowering of the scale $\Lambda$.
  
  But still an efficient calculation of these integrals requires adaptive routines.
  More precisely, the one-dimensional momentum integral in Eq.~(\ref{eq:num-int-ab0}) is then calculated using the routine \texttt{d01sjc} of the NAG library. \cite{nag}
  Conversely, we use the multidimensional quadrature routine \texttt{dcuhre} \cite{dcuhre} 
  for the two-dimensional integrals of Eqs.~(\ref{eq:num-int-yuk},\ref{eq:num-int-af0},\ref{eq:num-int-afx}).
  The infinite-range frequency integrals in these equations can be safely cut off at $ K_\tau = \ln 10^{13} $, as
  a variation of this upper cutoff around that value that does not affect the results within the given error tolerances.

  In order to reach the desired precision/accuracy goal, an ODE solver with a step width chosen according to these error tolerances is needed.
  We therefore use the \texttt{rksuite} code of Ref.~\onlinecite{rksuite}, where we choose a fourth-order Runge-Kutta algorithm with a fifth-order error estimate for the
  adjustment of the step width.
  The flow is solved by iteratively calling the \emph{complicated task} routine of \texttt{rksuite} for each Runge-Kutta step. After each step, the corresponding couplings 
  and associated anomalous dimensions are written to an output file. These anomalous dimensions are easily obtained, since
  \texttt{rksuite} provides estimates for the scale derivatives at each step.

  In order to obtain reliable results for the scaling exponents, we have set the precision goal for the Runge-Kutta solver to $10^{-7}$.
  The precision goal for the quadrature routines is then set to $10^{-8}$ in order to allow for a stable solution of the RG flow.
  The data points generated by \texttt{rksuite} are then relatively dense so that no interpolation is needed.

  We stress that these requirements can only be met through the usage of adaptive quadrature routines 
  and the logarithmic transformation of integration variables,
  at least for zero bosonic mass. However, we have found the flows with a fine-tuned bosonic mass 
  to be intractable with the numerical methods described here.

\bibliography{20151008-QLSM-references-corr}

\end{document}